%% file: main.tex
\documentclass[10pt,aps,prx,twocolumn,english,superscriptaddress,floatfix,longbibliography,nofootinbib]{revtex4-2}

\usepackage[table]{xcolor}
\usepackage[utf8]{inputenc}
\usepackage[english]{babel}

\usepackage{enumitem}

\usepackage{hyperref}
\hypersetup{
    unicode=true, 
    colorlinks=true, 
    linkcolor=teal, 
    citecolor=teal,
    urlcolor=teal
}

\usepackage[table]{xcolor}
\usepackage{tcolorbox}
\definecolor{light_blue}{HTML}{f0f5ff}
\definecolor{light_grey}{HTML}{ededed}
\tcbset{fontupper=\normalsize,
	colback=white, colframe=black, colbacktitle=light_grey,
	coltitle= black, arc=0.25mm}
\usepackage{latexsym,bm,amsfonts,amstext,graphicx,bbm,relsize,dsfont,lipsum,bold-extra,amsmath,amssymb,enumerate,float}
\usepackage[bb=dsserif]{mathalpha}
\usepackage{silence}
\usepackage{times}

\input{commands}

\newcommand{\Sdagger}[1]{\mathcal{S}^\dagger_{#1}}
\renewcommand{\T}[2]{\mathcal{T}^{\phantom{\dagger}}_{#1#2}}
\newcommand{\Tdagger}[2]{\mathcal{T}^\dagger_{#1#2}}

\begin{document}
\sloppy

\title{Optimisation of ultrafast singlet fission in 1D rings towards unit efficiency}

\author{Francesco Campaioli}
\email{francesco.campaioli@unipd.it}
\affiliation{Dipartimento di Fisica e Astronomia “G. Galilei” Università degli Studi di Padova, I-35131 Padua, Italy}
\affiliation{Padua Quantum Technologies
Research Center, Università degli Studi di Padova, Italy I-35131, Padova, Italy}
\affiliation{INFN, Sezione di Padova, via Marzolo 8, I-35131, Padova, Italy}

\author{Alice Pagano}
\affiliation{Institute for Complex Quantum Systems, Ulm University, Albert-Einstein-Allee 11, 89069 Ulm, Germany}
\affiliation{Dipartimento di Fisica e Astronomia “G. Galilei” Università degli Studi di Padova, I-35131 Padua, Italy}
\affiliation{Padua Quantum Technologies
Research Center, Università degli Studi di Padova, Italy I-35131, Padova, Italy}
\affiliation{INFN, Sezione di Padova, via Marzolo 8, I-35131, Padova, Italy}

\author{Daniel Jaschke}
\affiliation{Institute for Complex Quantum Systems, Ulm University, Albert-Einstein-Allee 11, 89069 Ulm, Germany}
\affiliation{Dipartimento di Fisica e Astronomia “G. Galilei” Università degli Studi di Padova, I-35131 Padua, Italy}
\affiliation{Padua Quantum Technologies
Research Center, Università degli Studi di Padova, Italy I-35131, Padova, Italy}
\affiliation{INFN, Sezione di Padova, via Marzolo 8, I-35131, Padova, Italy}

\author{Simone Montangero}
\affiliation{Institute for Complex Quantum Systems, Ulm University, Albert-Einstein-Allee 11, 89069 Ulm, Germany}
\affiliation{Dipartimento di Fisica e Astronomia “G. Galilei” Università degli Studi di Padova, I-35131 Padua, Italy}
\affiliation{Padua Quantum Technologies
Research Center, Università degli Studi di Padova, Italy I-35131, Padova, Italy}
\affiliation{INFN, Sezione di Padova, via Marzolo 8, I-35131, Padova, Italy}

\date{\today}

\begin{abstract} 
Singlet fission (SF) is an electronic transition that in the last decade has been under the spotlight for its applications in optoelectronics, from photovoltaics to spintronics. Despite considerable experimental and theoretical advancements, optimising SF in materials like multichromophoric systems and molecular crystals remains a challenge, due to the complexity of its analysis beyond perturbative methods.
Here, we tackle the case of 1D rings,
aiming to promote singlet fission and prevent its back-reaction. We study ultrafast SF non-perturbatively, by numerically solving a spin-boson model, via exact propagation and tensor network methods. By optimising over a parameter space relevant to organic molecular materials, we identify two classes of solutions that can take SF efficiency beyond 85\% in the non-dissipative (coherent) regime, and to 99\% when exciton-phonon interactions can be tuned. 
After discussing the experimental feasibility of the optimised solutions, we conclude by proposing that this approach can be extended to a wider class of optoelectronic optimisation problems.
\end{abstract}

\maketitle
\makeatletter

\begin{figure}[t]
    \centering
    \includegraphics[width=0.44\textwidth]{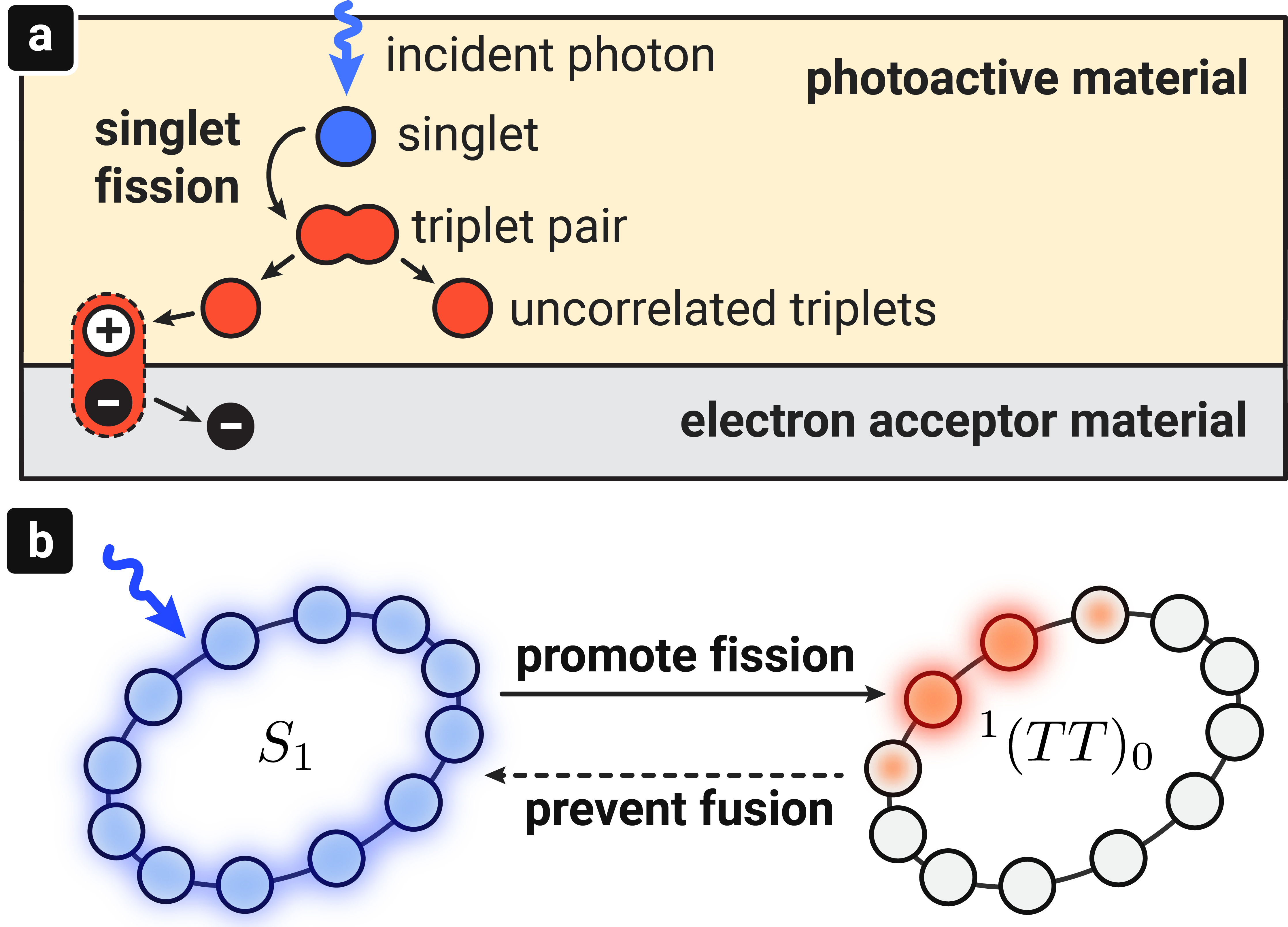}
    \caption{\textbf{Singlet fission in organic semiconductors.} Singlet fission (SF), a multi-exciton generation process in which two excitons are generated per absorbed photon, is the object of an intense interdisciplinary research effort due to its application in optoelectronics~\cite{lee2013singlet,xia2017singlet,tayebjee2017quintet,baldacchino2022singlet,carrod2022recent}. (a) For photovoltaics applications, singlets are formed via photoexcitation in a photoactive material and then split into triplet pairs. Triplets can migrate in the medium until they reach an electron acceptor material, where they can dissociate into charge carriers to produce a photocurrent. (b) Sketch of  the initial transient of SF in 1D rings, where a delocalised $S_1$ singlet splits into a triplet pair ${}^1(TT)_0$. Here, we aim to promote the generation of triplet pairs while discouraging the back-reaction, known as \textit{triplet fusion}~\cite{Atkins1975,Iwasaki2001,Gholizadeh2020,alves2022challenges,Forecast2023a,Forecast2023}.}
    \label{fig:1}
\end{figure}

In the last decade, \textit{organic optoelectronics}~\cite{Ostroverkhova2016}, i.e., the branch of electronics that focuses on light-matter interaction in organic semiconductors, has opened to a new generation of applications and perspectives in organic solar cells~\cite{Chenu2015,duan2020progress,fukuda2020future}, LEDs~\cite{gelinas2014ultrafast,song2020organic,sun2022exceptionally,zou2020recent}, low-light sensors~\cite{leung2014light,huang2015photostable,delPino2018}, magnetometry~\cite{budker2007optical,rizal2021magnetophotonics,matsko2005magnetometer}, microelectronics~\cite{high2007exciton,huo2014novel,sun2015single,cho2021recent,kumar2023optoelectronic}, and more~\cite{Ostroverkhova2016}. 
Singlet fission (SF) is one such optoelectronic process that has seen a surge in interdisciplinary studies~\cite{Tamura2015,Nakano2016,Ito2016,Nagashima2018,Kim2019,Schnedermann2019,Deng2019,Pun2019,Bayliss2020,Wollscheid2020,Korovina2020,Kim2021a,Dvorak2021,Jiang2021,Tsuneda2022,Collins2023,Neef2023}. It consists in the splitting of an electronic excitation (\textit{exciton}) with spin 0, known as a \textit{singlet}, into two excitons with spin 1, thus referred to as a \textit{triplet pair}~\cite{Berkelbach2013}, as illustrated in Fig.~\ref{fig:1} (a). This process, discussed formally in Sec.~\ref{s:theory}, has received lots of attention for its potential application in photovoltaics~\cite{lee2013singlet,xia2017singlet,tayebjee2017quintet,baldacchino2022singlet,carrod2022recent}, since multi-exciton generation---i.e., the generation of two or more excitons per absorbed photon---can lead to sizeable improvements in power conversion efficiency~\cite{kunzmann2018singlet,Tayebjee2015,baldacchino2022singlet}. Furthermore, triplet-mediated exciton transport can also bring benefits to organic solar cells by countering radiative emission losses that affect singlet excitons~\cite{snoke2002long,mullenbach2017probing,Davidson2022}. SF is also being applied in spintronics and information processing~\cite{wan2018transport,bayliss2020probing,smyser2020singlet}, excitonic logic~\cite{hudson2024framework}, sensors~\cite{nagata2018exploiting}, 3D printing~\cite{limberg2022triplet,sanders2022triplet,wong2023triplet}, and phototherapy~\cite{huang2018near,wei2021oxygen,liu2024achieving}.

The mechanism underlying SF is fairly well understood in weakly-correlated materials that allow for semiclassical and perturbative treatment~\cite{Berkelbach2013,Berkelbach2014,Casanova2018,Shushin2019,Schmidt2019}, such as diluted solutions~\cite{walker2013singlet} and low-density crystals~\cite{Felter2019b}, as well as small strongly-correlated compounds~\cite{Pun2019}, such as bridged molecular dimers~\cite{Aryanpour2015,Basel2017,Duan2020,Kim2021a,Schnedermann2019,Schroder2019,Tonami2023}. Indeed, design guidelines for optimal SF in these materials are now becoming available~\cite{Ito2016,Kumarasamy2017,Japahuge2019,Jacobberger2022,Collins2023}. However, studying SF in materials like multichromophoric systems and molecular crystals---a key class of materials for optoelectronics---remains a formidable challenge because of exciton delocalisation, entanglement, and the interplay between electronic and vibrational degrees of freedom~\cite{Xie2019,Miyata2019,Mukherjee2023}. Therefore, finding design principles for optimal SF in extended materials is a major outstanding challenge. 

Here, we build on recent developments~\cite{Teichen2015,delPino2018,Schnedermann2019,Schroder2019}, and optimise the initial ultrafast\footnote{The first picoseconds after photon absorption.} transient of SF in 1D extended materials, aiming to maximise the production of triplet pairs while preventing their recombination into a singlet.
In particular, we focus on 1D ring materials, illustrated in Fig.~\ref{fig:1} (b). These structures have emerged in plants and bacteria due to their performance as light-harvesting antennas~\cite{Jang2018}, which stems from a scalable enhancement of optical dipole moments, mediated by exciton delocalisation, that grows with the number $N$ of sites composing the medium~\cite{Hestand2018}, as illustrated in Fig.~\ref{fig:2} (a).
Interestingly, our results indicate that also SF benefits from the scaling of the system's size $N$, as we discuss in Sec.~\ref{s:results_discussion}. 

While singlet fission is typically studied perturbatively in the singlet-triplet interaction~\cite{Casanova2018} (here represented by the parameter $\gamma$),
recent experiments have shown that perturbation theory is insufficient to explain SF rates and steady-state populations already for simple 2-molecule compounds~\cite{Schnedermann2019,Schroder2019}. Therefore, here, we model singlet fission using a well-known spin-boson model~\cite{Teichen2015,Berkelbach2013,Schroder2019,Schnedermann2019} and solve non-perturbatively in $\gamma$ via exact propagation and tensor network methods (TNMs), up to $N = 128$.

We perform optimisation using standard routines based on a notion of efficiency that is inherited from a conserved quantity of the system's Hamiltonian. We find two promising classes of solutions that take SF efficiency, i.e., the probability to convert $n$ initial singlets into $2n$ triplets, (1) above 85\% in the non-dissipative regime (coherent, unitary) where excitons do not interact with vibrational phononic modes, and (2) to 99\% in the dissipative case where exciton-phonon interactions are considered. Our results for coherent SF indicate that disorder can have a beneficial role, which we compare and contrast with the well-known decoherence-enhanced exciton transport phenomenon in organic semiconductors~\cite{Mohseni2008,Plenio2008,Rebentrost2009,Kassal2013,Jeske2015,Jang2018,Mattioni2021}.
After discussing the experimental feasibility of our solutions, we conclude by arguing that quantum simulations and classical ``quantum-inspired'' ones offer a promising avenue to tackle a wide class of material optimisations in optoelectronics, with singlet fission being a paradigmatic case study for applying this approach. 

\section{Ultrafast singlet fission in 1D rings}
\label{s:theory}
\begin{figure}[t]
    \centering
    \includegraphics[width=0.47\textwidth]{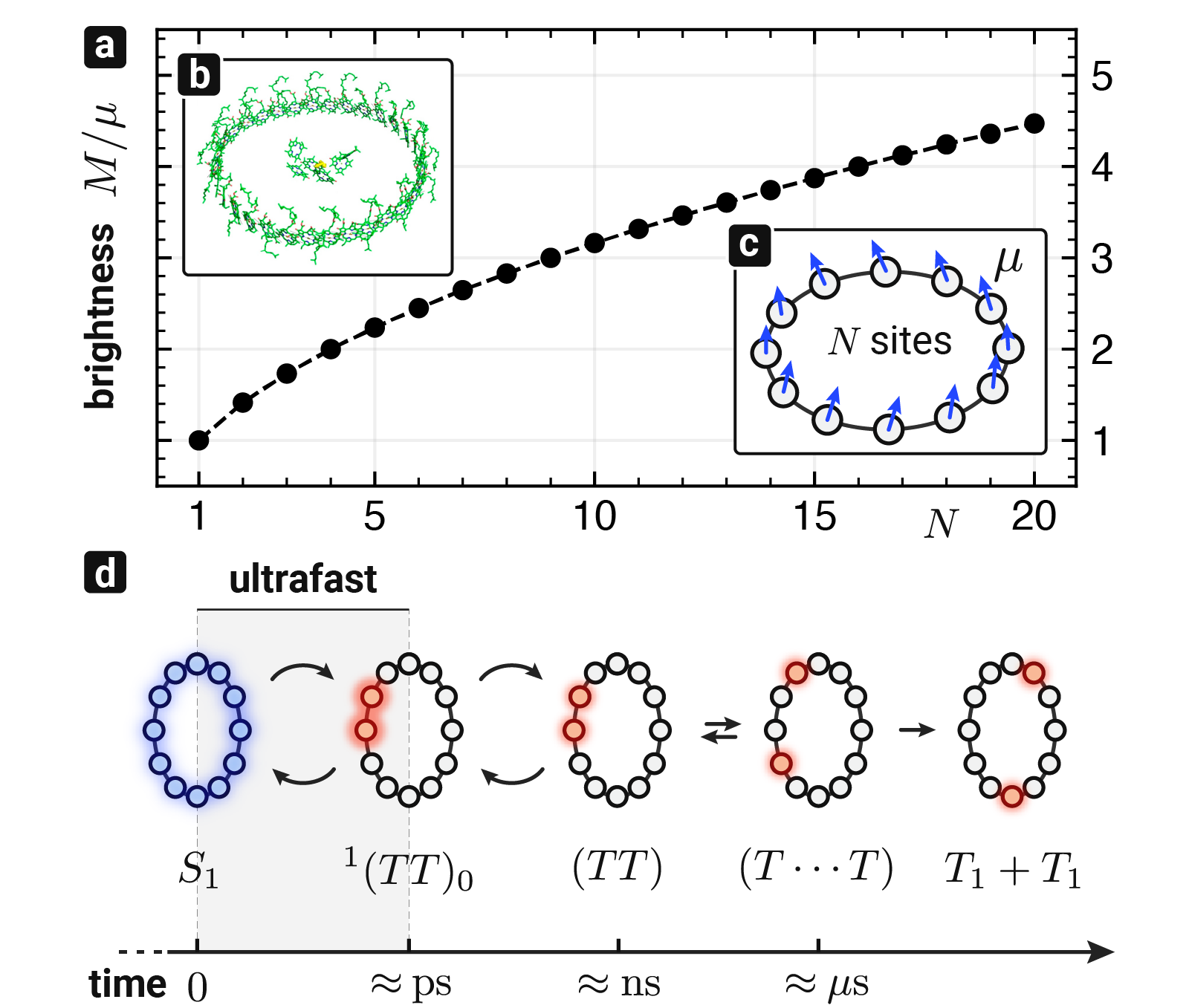}
    \caption{\textbf{Excitons in 1D rings.} (a) Excitons in 1D rings can have an enhanced total optical dipole moment $M$ that is $\sqrt{N}$ larger than the local dipole moment of each component $\mu$, due to delocalisation~\cite{baghbanzadeh2016distinguishing,baghbanzadeh2016geometry,tomasi2020classification,tomasi2021environmentally}. (b) 1D molecular rings have emerged in nature as exceptional optical antenna. The light-harvesting complex (LH1) of purple bacteria~\cite{scholes2010quantum} shown in the inset is a ring of $N\approx32$ chromophores that are coupled to a common exciton acceptor placed in the centre of the ring, known as the reaction centre (adapted from Ref.~\cite{pishchalnikov2021relationship}). (c) Here, we consider 1D rings composed of $N$ sites with dipoles of identical magnitude $\mu$ and varying orientations. (d) Sketch of SF across different timescales. We focus on the ultrafast SF transient that occurs immediately after photoexcitation, in which a singlet $S_1$ splits into a triplet pair with vanishing total spin ${}^1(TT)_0$. Over this timescale, exciton dynamics is partially coherent and non-Markovian due to exciton-phonon interactions~\cite{Schroder2019}. By interacting with the environment, a triplet pair can then acquire higher total spin, $(TT)$, separate across the medium, $(T\dots T)$, and eventually lose all correlations, $T_1 +T_1$~\cite{Berkelbach2013,Marcus2020,manian2023charge}. }
    \label{fig:2}
\end{figure}

In singlet fission, a spin-0 singlet exciton ($S_1$) splits into two spin-1 triplet excitons ($T_1$). The overall transition, summarised in Fig.~\ref{fig:2} (d) for the specific case of 1D rings (1D chains of $N$ sites with nearest-neighbour couplings and periodic boundary conditions), occurs across different timescales~\cite{Lee2018}. Here, we focus on the ultrafast transient $S_1 \rightleftharpoons {}^1(TT)_0$, that begins immediately after photoexcitation, and takes place in the femtosecond--picosecond timescales\footnote{Accordingly, we neglect all the processes that occur over much longer time scales, such as singlet and triplet exciton recombination, which occur in the nanosecond and microsecond timescales, respectively. We have also neglected other competing processes like triplet-pair annihilation to the ground state.}~\cite{Zheng2016}, where ${}^1(TT)_0$ is the triplet-pair state with vanishing total spin~\cite{Collins2019}. This transition occurs via short-range singlet-triplet interactions that couple $S_1$ and ${}^1(TT)_0$ (directly or indirectly via virtual population of a charge-separated state~\cite{Yao2015,Nakano2016,Kim2019,manian2023charge}). The efficiency of multi-exciton generation largely depends on this ultrafast transient~\cite{Miyata2019}. Since its timescale is comparable with that of exciton decoherence (which is mediated by exciton-phonon interactions) its simulation requires a full quantum mechanical treatment~\cite{Balzer2021,Schnedermann2019}, based on the solution of the time-dependent Schr\"odinger equation or a quantum master equation to the relevant electronic and vibrational degrees of freedom~\cite{Yalouz2017}. 

In this work we focus on organic optoelectronic materials, therefore we only consider Frenkel excitons, i.e., excited states of electron-hole pairs such that electron and hole are localised within the same site~\cite{schroter2015exciton,bardeen2014structure}, e.g., molecule. This approach allows us to model SF using a coarse-grained local basis for the $i$-th site as
\begin{equation}
    \label{eq:local_basis}
    \mathcal{B}_i:=\{\ket{S_0}_i,\ket{S_1}_i,\ket{T_{1}}_i\}, \;\;\; i=1,\dots,N,
\end{equation}
consisting of the ground singlet exciton $\ket{S_0}_i$, the first excited singlet exciton $\ket{S_1}_i$, and the triplet exciton $\ket{T_{1}}$. By choosing this basis we also neglect the possibility of having two or more excitons on the same site, which is motivated by the fact that doubly excited sites are typically out of the energetic range considered for SF~\cite{Miyata2019}. Note that we have implicitly collapsed the 3-dimensional triplet manifold into a unique state $\ket{T_1}$, as often done for these calculations~\cite{Teichen2015,Casanova2018,Miyata2019}. This is equivalent to limiting the dynamics to the triplet-pair states ${}^1(TT)_0$ with vanishing total spin ($S^2 = 0$, $S_z=0$), often known as triplet-pair states with ``singlet character''. This basis is valid at zero magnetic field and in the absence of zero-field splitting interactions. The latter can be accounted for by keeping track of the $m$ quantum number the local triplet states $\ket{T_1} \to \ket{T_{1,m}}$ (see Sec.~\ref{a:spinful_model} of the Appendix), with no effect on the presented results.

\subsection{Hamiltonian}
\label{ss:hamiltonian}

To study $S_1 \rightleftharpoons {}^1(TT)_0$, we use a well-known model based on an exciton-phonon Hamiltonian~\cite{Schnedermann2019},
\begin{equation}
    \label{eq:full_hamiltonian}
    H = H_\mathrm{ex} + H_\mathrm{ph} + H_\mathrm{ex-ph},
\end{equation}
which allows us to account for the interplay between excitons and phonons, i.e., the excitations of the vibrational degrees of freedom of the molecules composing the medium. The exciton Hamiltonian $H_\mathrm{ex}  = H_S + H_T + H_\mathrm{int}$ is given by a singlet term, a triplet term, and a singlet-triplet interaction~\cite{Teichen2015}. The singlet term consists in a local energy ($\varepsilon_S$) and a hopping interaction ($J_S$),
\begin{equation}
    \label{eq:singlet_term}
    H_S = \sum_{i=1}^N \varepsilon_S \mathcal{S}_i^\dagger \mathcal{S}_i^\pdagger + \sum_{i=1}^N \bigg( J_S \mathcal{S}_i^\dagger\mathcal{S}_{i+1}^\pdagger + h.c. \bigg),
\end{equation}
where $\mathcal{S}_i^\dagger$ is the singlet creation operator associated with the transition $\ket{S_0}_i \to \ket{S_1}_i$ between the local ground state $S_0$ and singlet excited state $S_1$ at site $i$. Note that the indices $(i,i+1) \in \{(1,2),(2,3), \cdots, (N,1)\}$ run over all the nearest-neighbour pairs with periodic boundary conditions. Similarly, the triplet Hamiltonian $H_T$ consists of a local energy ($\varepsilon_T$) and a hopping term ($J_T$), in addition to a triplet-triplet \textit{exchange} interaction ($\chi$), which here takes the form of a \textit{density-density} coupling\footnote{Note that density-density coupling interaction strength $\chi$ is often negative for triplet-pair states with singlet characters~\cite{Miyata2019}}.,
\begin{equation}
    \label{eq:triplet}
    \begin{split}
         H_T = &\sum_{i=1}^N \varepsilon_T \mathcal{T}^\dagger_i\mathcal{T}^\pdagger_{i} + \sum_{i=1}^N \bigg( J_T \mathcal{T}_i^\dagger\mathcal{T}_{i+1}^\pdagger + h.c. \bigg) + \\
        &\sum_{i=1}^N \chi \mathcal{T}_i^\dagger \mathcal{T}_{i+1}^\dagger \mathcal{T}_{i+1}^\pdagger \mathcal{T}_i^\pdagger,
    \end{split}
\end{equation}
where $\mathcal{T}_i^\dagger$ is the triplet creation operator associated with the transition $\ket{S_0}_i \to \ket{T_1}_i$. 

The singlet-triplet interaction $\gamma$ is treated phenomenologically~\cite{Teichen2015}, without making any assumption on its nature\footnote{This coupling can be direct or mediated by virtual population of charge transfer states~\cite{Miyata2019}.}, other than implying that it couples a pair of neighbouring triplets to a singlet exciton,
\begin{equation}
    \label{eq:singlet-triplet}
    H_\mathrm{int} = \sum_{i=1}^N \gamma \bigg( \mathcal{T}^\dagger_i\mathcal{T}^\dagger_{i+1}\mathcal{S}_i^\pdagger + \mathcal{T}^\dagger_i\mathcal{T}^\dagger_{i+1}\mathcal{S}_{i+1}^\pdagger + h.c.  \bigg).
\end{equation}
As mentioned earlier, singlet fission is often studied perturbatively in $\gamma$, using Fermi's golden rule (FGR), Green's function (GF) method, as well as Marcus and Bloch-Redfield theory~\cite{Berkelbach2013,Casanova2018}, by evaluating the transition rates between some eigenstates of $H_S$ and of $H_T$. A semi-analytical solution to this problem using a perturbative approach in $\gamma$ can be found in Sec.~\ref{a:perturbative_solution} of the Appendix. However, recent theoretical and experimental results have highlighted the limitations of perturbation theory even for small systems like conjugated dimers~\cite{Schnedermann2019,Schroder2019}. Therefore, we adopt a non-perturbative approach, showing that it is necessary to correctly capture the dependence of singlet fission efficiency on $\gamma$.

\begin{figure*}[t]
    \centering
    \includegraphics[width=0.98\textwidth]{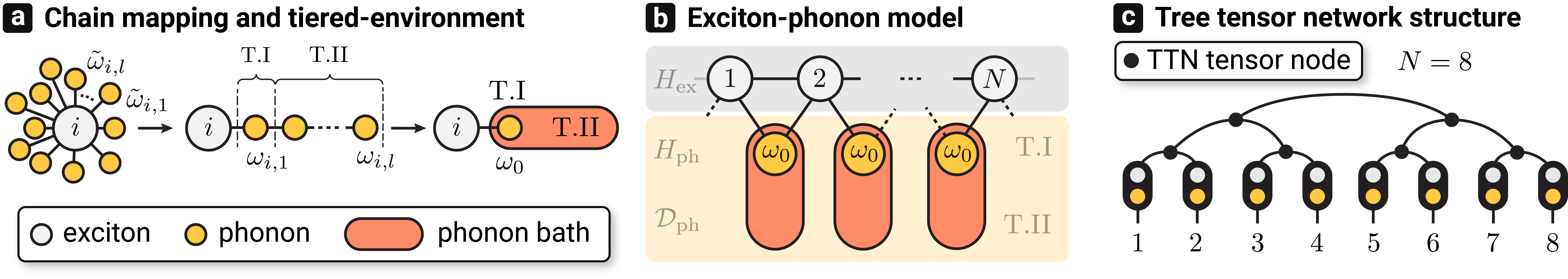}
    \caption{\textbf{Singlet fission model.} (a) Excitons couple to a large local ensemble of harmonic modes that represent the molecular vibrations~\cite{Miyata2019}. The chain mapping can be used to efficiently represent these harmonic modes as a 1D chain of quantum harmonic oscillators coupled to each excitonic site $i$~\cite{delPino2018,Schnedermann2019}. Using the tiered-environment approach~\cite{Fruchtman2015}, we group the first node of these 1D chains for every site $i$ as the first tier of the environment ($\mathrm{T.I}$). All the other nodes are grouped into a local bath of phonons ($\mathrm{T.II}$) that induce relaxation on the $\mathrm{T.I}$ nodes. (b) The exciton Hamiltonian $H_\mathrm{ex}$ is coupled to first tier ($\mathrm{T.I}$) of vibrational modes $H_\mathrm{ph}$ (with frequency $\omega_0$ and creation operator $a_i^\dagger$) linearly in the displacement operator ($a_i^\dagger + a_i^\pdagger$) as shown in Eqs.~\eqref{eq:phonons},~\eqref{eq:exciton-phonons}. The other vibrational modes belong to the second tier ($\mathrm{T.II}$) and are modelled as a Markovian bath that induces relaxation by acting directly only on $\mathrm{T.I}$ via the GKSL dissipator $\mathcal{D}_\mathrm{ph}$ given in Eq.~\eqref{eq:Lindblad}. (c) When modelling the system with TNMs, we used tree tensor network (TTN) with $U(1)$ symmetry in which each site is given by the embedding of an excitonic site (with local dimension $d_\mathrm{ex}$) and a T.1 phononic site (with tunable local dimension $d_\mathrm{ph}$), here represented for $N=8$. Due to the binary tree structure of the network we use TTNs for $N = 2^\kappa$, up to $N=128$ when phonons are ignored, and up to $N=8$ otherwise.}
    \label{fig:3}
\end{figure*}

We begin by considering coherent ultrafast SF~\cite{Miyata2019} in Sec.~\ref{ss:non-dissipative}, i.e., neglecting the effect of vibrational modes, as done in Ref.~\cite{Teichen2015}. This regime is relevant when exciton-phonon interactions occur at a much lower rate than SF. In Sec.~\ref{ss:dissipative}, we look at SF mediated by exciton-phonon couplings, which are known to play a key role in determining the exciton steady-state populations~\cite{Collins2023,Kobori2020}. Recent results~\cite{Fruchtman2015,Schroder2019,Schnedermann2019,delPino2018} model the vibrational modes as an ensemble of local, uncoupled harmonic oscillators that interact linearly with the excitons. This model has been extremely successful at simulating the dynamics of ultrafast singlet fission in molecular dimers~\cite{Schroder2019,Schnedermann2019} at the level of quantitative accuracy. Here, we treat the vibrational modes by directly adopting the well-known chain mapping~\cite{Prior2010, Chin2010, Chin2011, deVega2015,deVega2017,delPino2018,Schroder2019}, in which each local ensemble of vibrational modes is approximated with a 1D chain, not to be confused with the 1D ring representing the medium. Furthermore, we also employ the tiered-environment approach~\cite{Fruchtman2015,man2015harnessing,Collins2023} and keep track of the dynamics of the first node of each vibrational chain (tier-$\mathrm{I}$), treating all the other nodes as Markovian bath (tier-$\mathrm{II}$) weakly coupled with the first vibrational mode, as illustrated in Figs.~\ref{fig:3} (a) and (b). This leads us to the following effective phonon Hamiltonian,
\begin{equation}
    \label{eq:phonons}
    H_\mathrm{ph} = \sum_{i=1}^N \omega_0 a^\dagger_{i} a^\pdagger_{i},
\end{equation}
where $a_i^\dagger$, $a_i$ are the creation and annihilation operators of a harmonic oscillator with frequency $\omega_0$ at site $i$. The exciton-phonon interaction is
\begin{equation}
    \label{eq:exciton-phonons}
    H_\mathrm{ex-ph} = \sum_l\sum_{i=1}^N g_l A_{i}^{(l)} \otimes \Big(a^\dagger_{i} + a^\pdagger_{i} + x_{l,0} \Big),
\end{equation} $x_{l,0}$ is a scalar offset of the oscillator's displacement, and
where $A_{i}^{(l)}$ are exciton operators coupled vibrational mode at site $i$ with coupling strength $g_l$. These exciton coupling operators $A_{i}^{(l)}$ can involve at most two neighbouring sites\footnote{Long-range coupling between phonons have been studied for exciton transport and are known to play a role on decoherence and brightness of singlet excitons~\cite{Jeske2015}.}, as shown in Fig.~\ref{fig:3}, to allow for the vibrational modes to affect exciton-exciton interactions. Note that the index $l$ in Eq.~\eqref{eq:exciton-phonons} is summed over all the possible exciton coupling operators, as discussed later in Sec.~\ref{s:results_discussion}. When using tensor network methods (TNMs), we construct each site as the embedding of an excitonic site (with local dimension $d_\mathrm{ex}=3$) and tier-I phononic site (with tunable local dimension $d_\mathrm{ph}$), as shown in Fig.~\ref{fig:3} (c).

Note that to capture the properties of molecular materials we include disorder in the model, generalising each parameter in the exciton Hamiltonian by allowing for normally-distributed disorder, as in $\varepsilon_S \to \varepsilon_S^{(i)} = \varepsilon_S + \delta^{(i)}_{\varepsilon_S}$, where $\delta^{(i)}_{\varepsilon_S}$ is sampled from a normal distribution with zero average and $\sigma_{\varepsilon_S}$ standard deviation. This greatly extends the scope of the model beyond ordered 1D chains (treated perturbatively in Ref.~\cite{Teichen2015}). In Sec.~\ref{s:results_discussion} we discuss how disorder is pivotal in improving the efficiency of coherent ultrafast SF. While this effect is reminiscent of decoherence-enhanced transport~\cite{Mohseni2008,Plenio2008,Rebentrost2009}, the two differ in several aspects as we discuss in Sec.~\ref{ss:non-dissipative}.

\subsection{Initial state}
\label{ss:initial_state}

Photons are absorbed via excitation of singlet states, which, as opposed to triplets, have a strong optical dipole moment. Therefore, the initial exciton state is an eigenstate of the singlet Hamiltonian $H_S$ with a non-vanishing optical dipole moment~\cite{Teichen2015}. Note that the singlet Hamiltonian $H_S$ conserves the total number of singlets $\mathcal{N}_S$,
\begin{equation}
    \label{eq:total_number_singlets}
    \mathcal{N}_S:= \sum_i \mathcal{S}^\dagger_i \mathcal{S}_i^\pdagger,
\end{equation}
i.e., $[\mathcal{N}_S,H_S]=0$, inheriting the block structure
\begin{equation}
    \label{eq:singlet_block_structure}
    H_S=\sum_{n=0}^N H_S^{(n)},
\end{equation}
where $H_S^{(n)}$ are the $C(N,n) =\binom{N}{n}$-dimensional blocks associated with $n$ singlet excitons in the medium. 
We use this structure to select initial exciton states $\ket{\psi_0}$ with no triplets and $n_0 = \braket{\mathcal{N}_S}_0$ singlets. 

To select optically active states we use a notion of brightness that is based on the strength of the optical transition dipole moment of the state, assuming that each site $i$ has a dipole moment $\vec{\mu}_i$ with fixed magnitude $\|\vec{\mu}_i \|= \mu$ for all $i$. For $N<10$ we determine $\ket{\psi_0}$ exactly by diagonalising the sector $H_S^{(n_0)}$ and iterating over the eigenstates to find the one with the strongest dipole moment $M = \| \vec{M} \|$, where $\vec{M}$ is the total optical dipole moment vector associated with some eigenstate.
For $N\geq10$ we assume that $\ket{\psi_0}$ has the same structure. When using TNMs, we select $\ket{\psi_0}$ by computing the ground state of $H_S^{(n_0)}$ via density matrix renormalisation group (DMRG) on a tree tensor network (TTN) with $U(1)$ symmetry.

Note that, in the parameter region considered here for the singlet Hamiltonian, given by $\varepsilon_S/2>|J_S|$ and $\sigma_{\varepsilon_S}\ll J_S$, the brightest eigenstate with $n_0 = 1$ singlet is well approximated by the completely delocalised $W$-state\footnote{In analogy with the $W$-state, $\ket{W} = \frac{1}{N}(\ket{100\cdots0}+\ket{010\dots0}+\ket{0\cdots001})$ in quantum information theory~\cite{Mintert2005}.},
\begin{equation}
\label{eq:ground_singlet}
    \ket{W}:=\frac{1}{\sqrt{N}}\sum_{i=1}^N \mathcal{S}_i^\dagger\ket{S_0^{(N)}},
\end{equation}
which is the ground state of $H_S^{(1)}$,
where $\ket{S_0^{(N)}}:=\otimes_{i=1}^N\ket{S_0}_i$ is the collective ground state of the exciton Hamiltonian with no triplets nor singlets. It is important to mention that for $\ket{W}$ to be the brightest state, the dipoles need both a tangential component, i.e., along the ring in the ring's plane, and a transverse component, i.e., perpendicular to the ring's plane. For example, one can consider sites with dipole moments
\begin{equation}
    \label{eq:dipole_moment}
    \vec{\mu}_i = \bigg(-\mu_\circ\sin\bigg[\frac{(i-1)2\pi}{N}\bigg],\mu_\circ\cos\bigg[\frac{(i-1)2\pi}{N}\bigg],\mu_z\bigg),
\end{equation}
where $i$ is the site's index, $\mu_\circ$ is the strength of the tangential component and $\mu_z$ is the strength of the transverse component. By setting $\mu_z \leq \mu_\circ$ one obtains a bright $W$-state with $M = \sqrt{N}\mu_z$. Instead, if the dipoles have only a tangential component the $W$-state is dark since the vector sum of the molecular dipoles vanishes. Note that the choice of initial state can be easily adapted to different systems without affecting the performance of our optimisation approach.

When vibrational modes are also considered, the initial exciton-phonon state $\ket{\Psi_0}=\ket{\psi_0}\otimes\ket{\phi_0}$ is assumed to be the product between an initial exciton state $\ket{\psi_0}$ and the ground state of the vibrational modes before photoexcitation\footnote{In other words, the ground state of the vibrational modes assuming that the excitons are in the ground state $\otimes_{i=1}^N\ket{S_0}_i$.} $\ket{\phi_0}$,
\begin{equation}
    \label{eq:initial_state}
    \ket{\phi_0} = \otimes_{i=1}^N\ket{0}_i,
\end{equation}
where $\ket{0}_i$ is the ground state of the $i$-th harmonic oscillator $\omega_0 a_i^\dagger a_i^\pdagger$.
This corresponds to applying the \textit{Franck-Condon} principle~\cite{Deng2019,Sun2019}, meaning that the change in the electronic degrees of freedom induced by photon absorption is so sudden that the vibrational modes are out of equilibrium.

\begin{table*}
	\begin{tcolorbox}[tabulars*={\renewcommand\arraystretch{1.2}}%
	{l||c|c|c|c|c|c||c|c||c|c|c|c||c},adjusted title=flush left, halign title = left,
		boxrule=0.5pt,title = {\hspace{-8pt}\small\textbf{\textsf{Model parameters, range, and solution efficiency}}}]
		\hspace{2.5cm}  {} & $\varepsilon_S$ & $\varepsilon_T$ & $J_S$ & $J_T$ & $\chi$ & $\gamma$ & $\sigma_{J_T}$ & $\sigma_{\chi}$& $\omega_0$ & $x_0$ & $g_S$ & $g_T$ & $\eta$   \\
		\hline\hline
  {\textsf{\small{\textbf{Parameter range}}}} \\ 
  \hline
  {\textsf{\small{Minimum}}} & {$1$} & $0.35$ & $-0.2$ & $0$ & $0$ & $0.001$  & $0$ & $0$ & $0$ & $-0.1$ & $0$ & $0$ & {}   \\
  \hline
    {\textsf{\small{Maximum}}} & {$1$} & $0.65$ & $0.2$ & $0.5$ & $0.3$ & $0.6$ & $0.2$ & $0.2$ & $0.5$ & $0.1$ & $0.5$ & $0.5$ & {}   \\
		\hline\hline
  {\textsf{\small{\textbf{Solution}}}} \\ 
  \hline
  		{\textsf{\small{Resonant triplet-pair}}} & $1$ & {\small$\varepsilon_S/2-|J_S|$} & {\small $J_S<0$} & $0$ & $0$ & $\gamma$ & $0$ & $0$ & {-} & {-} & {-} & {-} & $0.50\pm0.35$  \\
		\hline
  {\textsf{\small{Optimised non-dissipative}}} & $1$ & $0.515$ & $-0.001$ & $0.3$ & $0.068$ & $0.437$ & $0.114$ & $0.005$ & {-} & {-} & {-} & {-} & $0.85\pm0.04$ \\
  \hline
 {\textsf{\small{Optimised dissipative}}} & $1$ & $0.372$ & $-0.001$ & $0$ & $0$ & $0.0103$ & $0$ & $0$ & $0.25$ & $-0.035$ & $0$ & $0.0038$ & $0.99\pm0.01$ \\
    \hline
	\end{tcolorbox}
 	\caption{\textbf{Model parameters, optimisation ranges, and SF efficiency.} The singlet energy is set to $\varepsilon_S = 1$ (in units of eV) for reference, with $\hbar\equiv 1$. The efficiency of the resonant triplet-pair solution is independent of the number of sites $N$ and singlet-triplet interaction $\gamma$, as long as the triplet energy $\varepsilon_T = \varepsilon_S/2-|J_S|$, where the singlet hopping coupling $J_T$ satisfies $-\varepsilon_S/2<J_S<0$. Other parameters are the triplet hopping coupling $J_T$, the triplet-triplet exchange coupling $\chi$, the phonons' frequency $\omega_0$, the phonons' oscillator offset $x_0$, the singlet-phonon coupling $g_S$, and triplet-phonon coupling $g_T$. Disorder in $J_T$ and $\chi$ is normally distributed with standard deviation $\sigma_{J_T}$ and $\sigma_\chi$, respectively.  The parameters' range allows for explorative optimisation while being representative of typical exciton and phonon energies and couplings in organic molecular materials~\cite{Ostroverkhova2016}.}
  \label{tab:solutions}
\end{table*}

\subsection{Dynamics}
\label{ss:dynamics}
To optimise SF, we study how well an initial singlet state is converted into triplets over a sufficiently long time. 
When the vibrational modes are ignored, we calculate the dynamics by evaluating or approximating,
\begin{equation}
    \label{eq:non-dissipative_dynamics}
    \begin{split}
        \ket{\psi(t)} &= U_\mathrm{ex}(t)\ket{\psi_0} ,\\&=e^{-i H_\mathrm{ex} t}\ket{\psi_0}.
    \end{split}
\end{equation}
Instead, when we consider exciton-phonon interaction, we propagate $\ket{\Psi_0}=\ket{\psi_0}\otimes\ket{\phi_0}$ using a quantum master equation for the dynamics of the composite exciton-phonon state. We use the Gorini–Kossakowski–Sudarshan–Lindblad (GKSL) master equation~\cite{breuer2002theory,Milz2017,campaioli2023tutorial} to the exciton-phonon density operator $\dot{\rho}_t = \mathcal{L}[\rho_t]$, 
\begin{align}
    \label{eq:Dissipator}
    \dot{\rho_t} &= -i[H,\rho_t] + \mathcal{D}_\mathrm{ph}[\rho_t], \\
    \label{eq:Lindblad}
    &= -i[H,\rho_t] + \sum^N_{\substack{{i=1} \\ k=\pm }} \gamma_k \bigg(L_{i,k}^\pdagger \rho_t L_{i,k}^\dagger - \frac{1}{2}\{L_{i,k}^\dagger L_{i,k}^\pdagger,\rho_t\}\bigg),
\end{align}
where the dissipator $\mathcal{D}_\mathrm{ph}$ only includes local transitions $L_{i,+} = a_i^\dagger$ and $L_{i,-} = a_i^\pdagger$, i.e., the local creation and annihilation operators, with associated transition rates $\gamma_+$ and $\gamma_-$, respectively.  
The latter is carried out using an exact propagation in Liouville space~\cite{campaioli2023tutorial} for $N\leq 4$ and a quantum-trajectory approach (Monte Carlo wavefunction method)~\cite{Carmichael1993,Daley2014,Donvil2022,jaschke2018one} combined with the time-dependent variational principle (TDVP) for TTNs with $U(1)$ symmetry~\cite{Haegeman2011,silvi2019tensor,Bauernfeind2020} for $N\geq4$. All the TNMs were implemented using the open-source library \href{https://pypi.org/project/qtealeaves/}{Quantum TEA}~\cite{qtealeaves_v0_5_12}. See Sec.~\ref{a:propagation} of the Appendix for details on the propagation methods. 

Before we present the optimisation results, let us notice that the total Hamiltonian has a conserved quantity $\mathcal{C}$ with $U(1)$ symmetry, such that $[H,\mathcal{C}] = 0$, 
\begin{equation}
    \label{eq:conserved_quantity}
    \mathcal{C} := 2\mathcal{N}_S + \mathcal{N}_T,
\end{equation}
that is also conserved by the dissipator $\mathcal{D}_\textrm{ph}$ of Eq.~\eqref{eq:Lindblad},
where $\mathcal{N}_T := \sum_{i=1}^N \mathcal{T}^\dagger_i \mathcal{T}_i^\pdagger$ is the total number of triplets. Since we assume that at $t=0$, i.e., immediately after photoexcitation, $\braket{\mathcal{N}_T}_0 = 0$, the number of triplets at time $t$ in our singlet fission event is bounded from above as 
\begin{equation}
    \label{eq:bound}
    \braket{\mathcal{N}_T}_t \leq \mathcal{C}_0 \equiv 2 \braket{\mathcal{N}_S}_0,
\end{equation}
i.e., twice the initial number of singlet excitons, where $\mathcal{C}_0:=\braket{\mathcal{C}}_0$. This conserved quantity allows us to study singlet fission dynamics within the sub-manifold defined by the initial value $\mathcal{C}_0$ of the conserved quantity. This limits the computational complexity of simulating SF to $\mathcal{O}(z^2)$, with $z = 2\binom{N}{\braket{\mathcal{N}_S}_0}$, which is exponentially hard as $\braket{\mathcal{N}_S}_0$ approaches half filling $N/2$. 
With TNMs this is handled by using quantum-number conserving (\textit{symmetric}) TTNs, thereby significantly reducing the computational cost of the simulation~\cite{montangero2018introduction,silvi2019tensor}. 

\section{Results and Discussion}
\label{s:results_discussion}
We now present the optimisation results.
To evaluate the performance of singlet fission we used a notion of \textit{efficiency} $\eta_t \in [0,1]$ that is inherited from the conserved quantity of Eq.~\eqref{eq:conserved_quantity},
\begin{equation}
    \label{eq:efficiency}
    \eta(t) := \frac{1}{2}\frac{\braket{\mathcal{N}_T}_t}{\braket{\mathcal{N}_S}_0},
\end{equation}
which serves as objective function. When studying SF in the dissipative regime, we evaluate $\eta(\tau)$ at sufficiently long times such that the number of triplets has reached a steady state.
Instead, to estimate the performance along individual trajectories and purely unitary dynamics without disorder we use the time-average efficiency $\overline{\eta}_\tau$ as objective function
\begin{equation}
    \label{eq:efficiency_time_average}
    \overline{\eta}_\tau:=\frac{1}{\tau}\int_0^\tau \eta(t)\:dt.
\end{equation}
In either case, we consider time intervals $\tau$ that range between $5\gamma^{-1}$ and $100\gamma^{-1}$. For reference, this amounts to about 100~fs to 5~ps when assuming $\gamma \approx 0.1\; \text{eV}$. 

Let us note that our choice of singlet fission efficiency is defined solely upon singlet and triplet populations (local observables), irrespective of other triplet-pair state properties such as their entanglement, classical correlations, and separation across the medium (non-local observables). While this is not a limitation on the validity of Eq.~\eqref{eq:efficiency}, these non-local observables are known to play a role in singlet fission efficiency. For example, the initial triplet-pair state is often characterised by bipartite entanglement~\cite{Teichen2015,HyungjunKim2018,Kim2024}, which can be lost over time due to several sources of decoherence such as nuclear spins, other electrons, other local and global magnetic fields~\cite{Marcus2020}, vibrational modes~\cite{manian2023charge}, and triplet transport across a disordered medium. In this work, we only discuss explicitly the role of triplet separation in Sec.~\ref{ss:non-dissipative} and in Sec.~\ref{a:triplet-separation} of the Appendix.

\subsection{Non-dissipative solution}
\label{ss:non-dissipative}
We begin by looking at the case of non-dissipative evolution, i.e., unitary (coherent) for the exciton, obtained by neglecting exciton-phonon interactions. This approximation is valid when exciton-phonon interactions do not occur on average within the considered time interval $\tau\approx\hbar\gamma^{-1}$ (picoseconds)~\cite{Collins2019,Alvertis2019}. In practice, this may be explored experimentally by working at low temperatures (7---80~K)~\cite{Pun2019,Tayebjee2016}, to limit the effect of noise induced by exciton-phonon interactions.

\begin{figure*}[ht]
    \centering
    \includegraphics[width=0.98\textwidth]{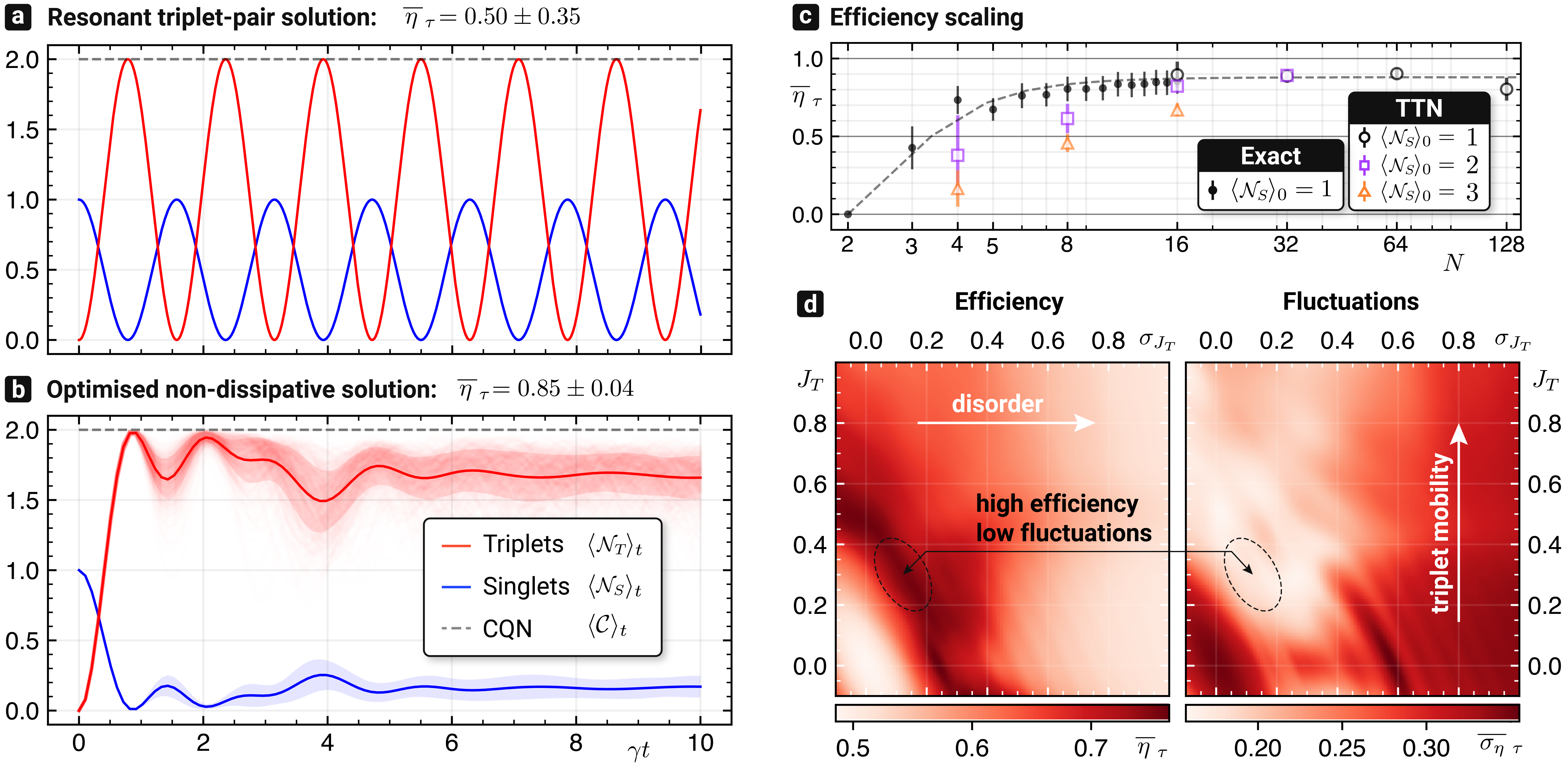}
    \caption{\textbf{Optimisation of non-dissipative ultrafast SF.} (a) The analytic \textit{resonant triplet-pair} solution ($\varepsilon_T=\varepsilon_S/2-|J_S|$, $J_T,\chi=0$) is here plotted for $\braket{\mathcal{N}_S}_0 = 1$ as a function of time ($\gamma t$) in units of the singlet-triplet interaction $\gamma$. Although it leads to the highest perturbative SF rate $\Gamma_\mathrm{SF} = 2\pi\gamma^2/\hbar$ for small $\gamma$, its time-averaged efficiency $\overline{\eta}_\tau = 0.50\pm 0.35$ is low, affected by large fluctuations, and independent of the system size $N$. (b) Optimised solution obtained for $N=10$, model parameters shown in Tab.~\ref{tab:solutions}, with efficiency $\overline{\eta}_\tau = 0.85\pm 0.04$, here shown by averaging over 500 disorder realisations. The solid lines are the average over the realisation, the shaded areas show the standard deviation around the average. The thin faint lines are the 500 trajectories for $\braket{\mathcal{N}_T}_t$. The conserved quantum number (CQN) $\braket{\mathcal{C}}_t \equiv \mathcal{C}_0$ is plotted for reference (dashed grey line). (c) The efficiency of the optimised solution grows as a function of the ring size $N$, and declines with the number of initial singlets $\braket{\mathcal{N}_S}_0 = 1,2,N/2$, since triplet fusion is promoted when the material is crowded with triplets. The efficiency of the optimised solution is at least 20\% higher than that of the resonant triplet-pair for $N\geq 4$. Error bars represent fluctuations around the average, solid (open) markers are calculated with ED (TTNs). The dashed lines fit to phenomenological model $\eta(N,n)=a(N-2)/\sqrt{(N-2)^2+b}$ with $a=0.87$ and $b=4$. The open markers represent results obtained using TTNs with bond dimension up to $100$. TTNs results are a lower-bound estimation of the efficiency for large $N$ due to the approximation imposed by limited bond dimension, as discussed in Sec.~\ref{a:propagation} of the Appendix. (d) A small amount of disorder $\sigma_{J_T}$ in the triplet hopping coupling $J_T$ is pivotal to increasing the efficiency and reducing fluctuations, as shown here for $N=6$.}
    \label{fig:4}
\end{figure*}

In this regime, we have an exact reference solution\footnote{We have also used this solution as a reference when testing the performance of the time-dependent variational principle (TDVP) propagation method used in combination with TTNs for large systems.}, given by the \textit{resonant triplet-pair} condition in ordered 1D rings~\cite{Teichen2015}. A resonant triplet pair forms when the singlet hopping coupling $J_S < 0$ satisfies $|J_S| < \varepsilon_S/2$, and the triplet energy $\varepsilon_T = \varepsilon_S/2-|J_S|$ is resonant with the initial delocalised singlet, for arbitrary $\gamma$ and vanishing triplet hopping coupling and triplet-triplet interaction $J_T, \chi = 0$. Similar solutions exist also for delocalised ($J_T\neq 0$) and interacting ($\chi \neq 0$) triplets. 
For these parameters, the system absorbs one photon with energy $\varepsilon_S-2|J_S|$ only via the perfectly delocalised 1-singlet state $\ket{W}$ of Eq.~\eqref{eq:ground_singlet}, which has a superradiant optical dipole moment $M = \sqrt{N}\mu_z$, leading to the enhanced absorption rate $\Gamma \propto N|\mu_z|^2$~\cite{Spano2010}. The unitary $U_t = \exp[-i H_\mathrm{ex} t]$ generated by the exciton Hamiltonian induces the formation of triplet pairs via the interaction term $H_\mathrm{int}$ of Eq.~\eqref{eq:singlet-triplet}, leading to perfect Rabi-like oscillations in the number of triplets
\begin{equation}
    \label{eq:resonant_solution}
    \braket{\mathcal{N}_T}_t = \cos\Big(4\gamma t + \pi\Big)+1,
\end{equation}
as shown in Fig.~\ref{fig:4} (a) for both triplets and singlets.

When studying SF perturbatively in $\gamma$ using FMG or GF methods~\cite{Teichen2015}, the \textit{resonant pair} solution is often considered ``optimal''\footnote{This is not generally the case when using Marcus theory or other perturbative theories that estimate transition rates mediated by a weakly coupled Markovian environment at thermal equilibrium.}, meaning that it leads to the largest singlet fission rate $\Gamma_\mathrm{SF} = 2\pi \gamma^2 /\hbar$. However, the back-reaction (the recombination of triplets into a singlet) happens at the same rate as SF, leading to 50\% efficiency in the long time average limit, $\lim_{\tau\to\infty} \overline{\eta}_\tau = \frac{1}{2}$, due to the detailed balance condition,
\begin{equation}
    \label{eq:detailed balance}
     \frac{1}{2}\frac{\braket{\mathcal{N}_T}_\infty}{\braket{\mathcal{N}_S}_\infty}= \frac{\Gamma_\mathrm{SF}}{\Gamma_\mathrm{TF}} \equiv 1,
\end{equation}
where $\Gamma_\mathrm{TF}$ is the rate of the back-reaction, known as triplet fusion (TF) or triplet-triplet annihilation.

Here, we aim to increase the efficiency $\eta$ and reduce its fluctuations\footnote{These can be calculated as the standard deviation of $\eta$ over the ensemble of disorder realisation, or as the time-averaged fluctuations around the $\overline{\eta}_\tau$.} $\sigma_\eta$.
To obtain an improved solution, we use a variety of optimisation methods\footnote{Some methods that we used were Sequential Least Squares Programming, Basing Hopping, and Nedler-Mead.} available in the Python's library \href{https://scipy.org/}{SciPy}, based on \href{https://docs.scipy.org/doc/scipy/reference/generated/scipy.optimize.minimize.html#scipy-optimize-minimize}{scipy.optimize.minimize} applied to the objective function $O:=-\overline{\eta}_\tau$. First, we carry out an initial rough optimisation using exact propagation by exploring a coarse-grained region of the full set of parameters given in Tab.~\ref{tab:solutions}, while keeping $N = 6$ and setting $\varepsilon_S = 1$. We then perform a finer optimisation using exact propagation for $N=10$ over the parameters $\varepsilon_T$, $J_S$, $J_T$, $\chi$,$\gamma$, as well as the disorder parameters $\sigma_{J_T}$ and $\sigma_{\chi}$. The efficiency of the solution reported in Tab.~\ref{tab:solutions} is well over 80\% for $N=10$, and improves as the system grows in size as shown in Fig.~\ref{fig:4} (b) and (c). Furthermore, in this regime the number of triplets tends to rapidly equilibrate~\cite{gogolin2016equilibration,Wilming2019}, leading to relatively small efficiency fluctuations both along a given trajectory and on average with respect to disorder realisations.

Let us now comment on the nature of such high SF efficiency. The region of parameter space around the optimal solution has the following properties:
\begin{itemize}[label={},leftmargin=0pt]
    \item \textbf{Significant singlet delocalisation}, activated by a weak but non-vanishing hopping $J_S \approx (-0.05,-0.001)$ in units of $\varepsilon_S$. By being delocalised, the singlets can ``see'' all the $\mathcal{O}(N)$ pairs of neighbouring sites where SF can occur, rather than just the two pairs associated with any given site. Additionally, these delocalised singlets also benefit from photoexcitation rate enhancement due to the superradiant optical dipole moment $M\propto \sqrt{N}\mu_z$.
    \item \textbf{Fast triplets}, with intermediate hopping $J_T \approx (0.1,0.3)$. Hopping promotes the fast separation of the triplets within the ring as soon as they are formed. Quantitative data on triplet-pair separation can be found in Sec.~\ref{a:triplet-separation} of the Appendix, where we show that our optimised solution is characterised by triplets that separate across distant chromophores. Conversely, in the resonant triplet-pair solution triplet-pairs are limited to adjacent chromophores.
    \item \textbf{Repulsive triplet-triplet interactions}, activated by $\chi \approx (0.02,0.08)$. These promote the separation of individual triplets and discourage their recombination into singlets at neighbouring pairs of sites.
    \item \textbf{Strong, resonant singlet-triplet coupling}, by setting $\gamma \approx 0.4$ and triplet energy $\varepsilon_T \approx \varepsilon_S/2$. This ensures that the triplets form rapidly and successfully before they can scramble ballistically across the ring.
    \item \textbf{Disorder} in the triplet hopping term $\sigma_{J_T} \approx 0.1$ and in the triplet-triplet interactions $\sigma_\chi \approx 0.005$. This promotes the ``scrambling'' of the state across the triplet manifold, discouraging the periodic re-population of the singlet states. The beneficial role of disorder is illustrated in Fig.~\ref{fig:4} (d), where it is evident that $\sigma_{J_T} = 0$ leads to large fluctuations and small SF efficiency. Note that while a moderate amount of disorder is beneficial, too much disorder is detrimental as it leads to triplet localisation (and thus slow triplet transport) and off-resonance between the singlet and the triplet pairs.
\end{itemize}

Note that the beneficial role of disorder highlighted in Fig.~\ref{fig:4} (d) is reminiscent of the well-known \textit{dephasing-assisted transport} effect~\cite{Mohseni2008,Plenio2008,Rebentrost2009,Sneyd2021,Campaioli2021}, but the two should not be confused. Dephasing-assisted transport is an open-system phenomenon mediated by weak interactions between the excitons and a Markovian bath of harmonic modes, that typically couple to the exciton site energy $\varepsilon_S$. Under some particular conditions, dephasing can aid transport by overcoming localisation, triggering a diffusive transport regime~\cite{Rebentrost2009}. Instead, the disordered-assisted enhancement of SF discussed here is a closed-system phenomenon triggered by random, normally distributed couplings in the triplet hopping and density-density interactions. Rather than being the result of decoherence, which comes with an entropic cost, disorder promotes a rapid isentropic equilibration process~\cite{gogolin2016equilibration}, that leads to a suppression of the fluctuations in the number $\braket{\mathcal{N}_T}_t$ of triplets.
This effect is novel in the context of SF and could be further explored to determine the extent of its benefits as a function of the system size, dimensionality, and magnitude of disorder.

We evaluated this solution for $N=16,32,64,128$ using TNMs with bond dimension up to $100$, while also varying the initial number of singlets $\braket{\mathcal{N}_S}_0 = 1,2,3$. Interestingly, our results show that 1D organic molecular rings with 32 sites (thus comparable with the LHC1 centre of purple bacteria~\cite{scholes2010quantum}) can host 3 simultaneous SF events at an efficiency that is close to 90\%. Intuitively, the efficiency is expected to decrease as $\braket{\mathcal{N}_S}_0/N$ increases, since the triplet-triplet encounters occur more frequently as the ring gets crowded with triplets. 
Indeed, an SF-blockade effect occurs whenever the ring is completely filled with singlet excitons, meaning that $\braket{\mathcal{N}_S}_0 = N$ leads to $\eta_t \equiv 0$. This also suggests that the efficiency of this solution would improve in 2D and 3D materials, where a larger number of configurations aids triplet pair separation.

\subsection{Dissipative solution}
\label{ss:dissipative}
It is well known that exciton-phonon interactions play a significant role in SF, which can be quite efficient (even above 80\%) when aided by vibrational modes~\cite{Nakano2016,Schnedermann2019,Schroder2019,Kumarasamy2017,Alvertis2019} for the case of molecular dimers. When singlet-triplet couplings $\gamma$ are weak and mediated by the environment, perturbative solutions (e.g., Marcus Theory) suggest the SF proceeds efficiently when the excess energy $Q_\mathrm{out} = \varepsilon_S - \varepsilon_{TT} > 0$ is dissipated into the environment as heat~\cite{Casanova2018}. Accordingly, as the efficiency $\eta$ of SF increases, its \textit{thermodynamic efficiency} $\eta_\mathrm{th} = 1 - {Q_\mathrm{out}}/{\varepsilon_S}$ tends to decrease. The latter eventually limits the overall power conversion efficiency in photovoltaics, as part of the energy of the absorbed photons is lost due to thermal dissipation rather than used to produce a photocurrent~\cite{Tayebjee2015}. 

Meanwhile, the nature of SF efficiency beyond perturbative solution is far less understood, especially for the vastly unexplored case of extended solids. For example, the puzzling case of \textit{exoergic}\footnote{Singlet fission where the resulting uncorrelated triplets $T_1+T_1$ have a larger energy than the initial singlet $S_1$~\cite{Miyata2019}. This can only happen if part of the energy necessary for the transition is provided by the vibrational environment.} SF is still object of an intense experimental and theoretical activity~\cite{Miyata2019, Pun2019}. Here, we study dissipative non-perturbative SF in a regime where exciton-phonon interactions occur over the $\hbar\gamma^{-1}$ timescale, i.e., picoseconds in organic molecular materials. We apply the optimisation approach discussed above to the full Hamiltonian of Eq.~\eqref{eq:full_hamiltonian} and master equation~\eqref{eq:Lindblad} to uncover some design guidelines for optimal dissipative SF, and discuss how to achieve near-unit efficiency even in the non-perturbative regime. Note that these calculations are far more demanding than those of Sec.~\ref{ss:non-dissipative}, since the local dimension scales as $d_\mathrm{ex}d_\mathrm{ph}$ due to the exciton-phonon embedding discussed in Fig.~\ref{fig:3} (c), which is why TNMs are necessary beyond $N=4$. For reference, the total Hilbert space dimension scales as that of $N_\mathrm{qbit} = N\log_2(d_\mathrm{ex}d_\mathrm{ph})$ qubits, which equates to around 200 qubits for $N=64$ and $d_\mathrm{ph}=3$.
\begin{figure}
    \centering
    \includegraphics[width=0.48\textwidth]{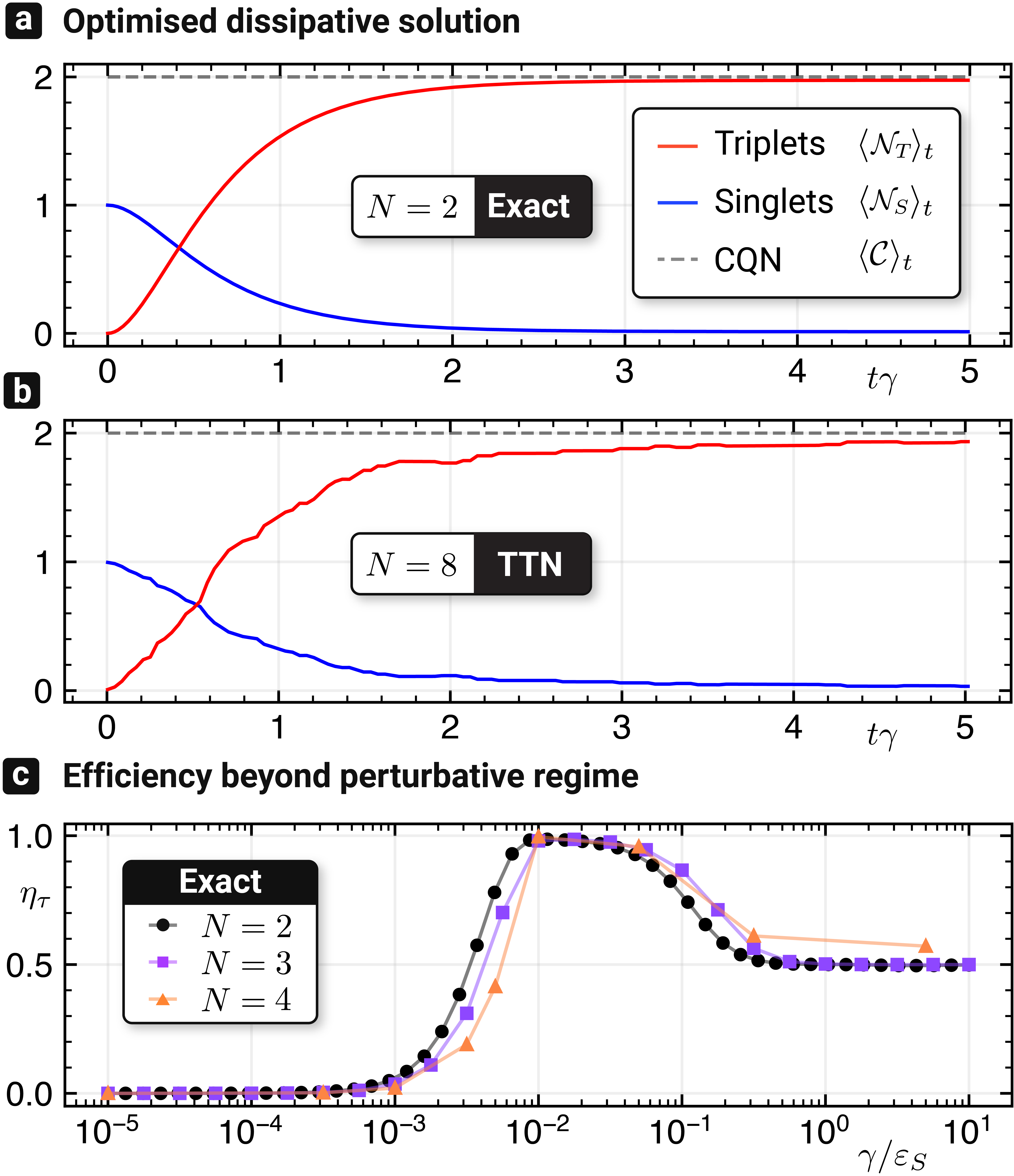}
    \caption{\textbf{Optimisation of dissipative ultrafast SF.} Exciton-phonon interactions can be tuned to achieve a perfect ultrafast singlet-to-triplet switch. The singlet and triplet densities are calculated using Eq.~\eqref{eq:Lindblad} for (a) $N=2$ using an exact propagation approach for the superoperator $\mathcal{L}$ and (b) for $N=8$ by simulating 100 quantum trajectories using the TDVP algorithm for quantum-number considering TTNs. (c) SF efficiency dependence on the singlet-triplet interaction $\gamma$. Perturbative solutions in $\gamma$ predict SF rates $\Gamma_\mathrm{SF}\propto\gamma^2$ and result in efficiencies that are constant in $\gamma$ and depend on the energy difference $\Delta E = \varepsilon_{S}-\varepsilon_{TT}$ between the initial singlet state and the triplet-pair states via the bath noise power spectrum $S(\Delta, k_\mathrm{B}T)$. The latter typically respects the thermodynamic detailed balance condition. 
    Instead, when the single-triplet coupling $\gamma$ is treated non-perturbatively, SF efficiency is non-monotonic in $\gamma$ and may not respect thermodynamic detailed balance condition.}
    \label{fig:5}
\end{figure}

In Fig.~\ref{fig:5} (a) we show the results that we obtained for $N=2,3,4,8$, by optimising over the exciton parameters discussed in Sec.~\ref{ss:non-dissipative} as well as the phonon frequency $\omega_0$, and the following exciton-phonon coupling parameters: The singlet-phonon coupling strength $g_S$, associated with the coupling terms
\begin{equation}
\label{eq:singlet_local_phonon}
    \sum_i {g_S} \mathcal{S}^\dagger_i\mathcal{S}^\pdagger_i\otimes \Big(a_i^\dagger+a_i^\pdagger\Big), 
\end{equation}
the triplet-phonon coupling strength $g_T$, associated with the coupling terms
\begin{equation}
    \label{eq:triplet_local_phonon}
    \sum_i g_T \mathcal{T}^\dagger_i\mathcal{T}^\pdagger_i\otimes \Big(a_i^\dagger+a_i^\pdagger\Big),
\end{equation}
and the singlet-triplet interaction strength $\gamma$, associated with the coupling term
\begin{equation}
    \label{eq:singlet-tirplet_phonon}
    \sum_{(i,j)} \gamma \bigg( \mathcal{T}^\dagger_i\mathcal{T}^\dagger_j\mathcal{S}_i + \mathcal{T}^\dagger_i\mathcal{T}^\dagger_j\mathcal{S}_j + h.c.  \bigg)\otimes\Big(a^\dagger_i+a_i^\pdagger + x_0\Big),
\end{equation}
where $x_0$ is a common scalar offset for the harmonic oscillators. Note that the latter
replaces the singlet-triplet interaction term $H_\mathrm{int}$ of Eq.~\eqref{eq:singlet-triplet}. The terms of Eqs.~\eqref{eq:singlet_local_phonon} and~\eqref{eq:triplet_local_phonon}  modulate the energy of the local singlet and triplet excitons, respectively. Note that the ratio between the relaxation rates $\gamma_\pm$ discussed below Eq.~\eqref{eq:Lindblad} is effectively a proxy for the temperature $T$ (or thermal energy $k_\mathrm{B} T$) of the bath, via $\gamma_+/\gamma_- \propto \exp(-\omega_0 / k_\mathrm{B} T)$. Here, we assume $\gamma_+ = 0$ since $\omega_0>0.1$ is significantly larger than thermal energy for any $T$ below room temperature ($k_\mathrm{B} T\approx 0.025$ at $T\approx 300~\mathrm{K}$). In particular, we fix $\gamma_- = 0.1$ and optimise over the other parameters.

As anticipated, we find optimal SF solutions in the dissipative regime that reach $\eta \geq 0.99$ and are therefore characterised by vanishing fluctuations $\sigma_\eta  \leq \approx 0.01$, reported in Tab.~\ref{tab:solutions}. In this solution, exciton-phonon interactions are tuned to perform a perfect $S_1\to{}^1 (TT)_0$ switch, shown in Fig.~\ref{fig:5} (a). This switch is achieved when singlets and triplets are coupled at the ground state configuration of the vibrational modes, i.e., immediately after photoexcitation, and not coupled at the equilibrium of the vibrational configuration of the triplet pair. Our solution presents an optimum in $\gamma$, illustrated in Fig.~\ref{fig:5} (b), in stark contrast with perturbative solutions. The latter predicts SF/TF rates that are both proportional to $\gamma^2$, leading to SF efficiency $\eta$ that is constant in $\gamma$ and that predominantly depends on the density of initial and final states and the bath noise power spectra. We also note that despite the high SF yield, our solution is characterised by a significantly lower thermodynamic efficiency $\eta_\mathrm{th} \approx 1-(\varepsilon_S-2\varepsilon_T)/\varepsilon_S \approx0.25$, due to the low triplet energy $\varepsilon_T < \varepsilon_S/2$. Nevertheless, we cannot rule out the existence of solutions that display a high SF yield without significant energetic losses, due the size of the parameter space and the complexity of the problem.

\section{Conclusions}
\label{s:conclusions}

In this work, we have optimised ultrafast singlet fission in 1D rings, finding two classes of solutions with near-perfect singlet-to-triplet conversion efficiency. We have shown that the efficiency of coherent, non-dissipative SF improves with the size $N$ of the system and benefits from a small amount of disorder. We have also shown that exciton-phonon couplings can be exploited to completely prevent the back reaction, by appropriately tuning the frequency $\omega_0$ of the vibrational modes and their coupling strength $\gamma$ to the singlet-triplet interaction.
This effectively corresponds to tuning the noise power spectrum of the vibrational modes that mediate singlet fission.

The optimisation has been performed by limiting the model parameter to a range that is representative of typical exciton/phonon energies and couplings in organic molecular semiconductors, which are a key class of materials for SF~\cite{Ostroverkhova2016}.
Nevertheless, the practical feasibility of these solutions depends on the controllability of the excitonic degrees of freedom, such as singlet and triplet exciton energies, optical dipole moments, and the relative arrangements between sites~\cite{Hestand2018}. The latter affect all exciton interactions, including singlet-triplet couplings. Recent developments in this direction have demonstrated controllability in most of these degrees of freedom~\cite{Meftahi2020a, dietrich2016exciton}. The tunability of exciton-phonon coupling is also becoming more feasible. For example, it has been shown that some vibrational modes can be suppressed or enhanced using functional groups in organic molecules~\cite{takeuchi2003raman,bandyopadhyay2014convenient,kurouski2013amide}. The associated vibrational noise power spectra can then be probed with Raman spectroscopy~\cite{horvath2016probing,myers2005resonance}. The fabrication and characterisation of these materials is an ongoing interdisciplinary challenge~\cite{chauhan2022organic}, which can benefit greatly from predictive tools like those used in this work.

In conclusion, our results are a key step towards finding design principles for optimal singlet fission in extended 2D and 3D solids. 
An important outlook in this direction is to explore efficient approaches to simulate SF beyond 1D systems. Tree Tensor Networks are currently used to tackle small 2D~\cite{jaschke2022ab,kshetrimayum2017simple} and 3D systems~\cite{Tepaske2021}. Neural network ansatz could prove valuable for the non-dissipative regime~\cite{vicentini2019variational,wu2023tensor}. Another promising avenue is to map these problems to equivalent spin-\textonehalf{} models that can be simulated on quantum platforms such as IBM Quantum~\cite{steffen2011quantum}, QuEra~\cite{ebadi2022quantum}, and PASQAL~\cite{chen2023continuous}, while taking advantage of environment-induced noise to simulate the effect of the vibrational modes~\cite{HQS2023}. Trapped-ion quantum simulations are already bringing insights into the coherent interplay between exciton and harmonic vibrational modes, which can be encoded in the normal modes of motion of the ion traps~\cite{Gambetta2021,whitlow2023quantum,so2024trapped}. This avenue could also be explored with quantum optimal control~\cite{ROSSIGNOLO2023108782} and machine learning optimisation methods~\cite{sutton2018reinforcement}.

Interestingly, the approach used here for SF can be extended to a wide class of electronic transitions that underlie the performance of many optoelectronic devices. For example, the efficiency of exciton transport~\cite{sneyd2022new}, widely studied using spin-boson models~\cite{Myers2018,Davidson2020,Sneyd2021}, has been identified as one of the limiting factors in organic solar cells~\cite{Kippelen2009}. Other examples include organic LEDs~\cite{raman2019generating}, low-light optical sensors~\cite{Lucas2014}, and molecular photo-switches used to functionalise RNA~\cite{mondal2015search}. A significant challenge is posed by the computational complexity of solving many-body spin-boson models, especially when symmetries like the one set by $\mathcal{C}$ in Eq.~\eqref{eq:conserved_quantity} are lacking. Another challenge is understanding the relationship between the microscopic parameters of the spin-boson model and the macroscopic performance indicators (e.g., the power conversion efficiency in photovoltaics) that determine the quality of a device. This challenge is highly cross-disciplinary and could bring a powerful approach to material and device optimisation.

\section*{DATA AND CODE AVAILABILITY}
The simulations were performed using the Quantum
Green Tea software version 0.3.26 and Quantum Tea
Leaves version 0.5.8~\cite{qtealeaves_v0_5_12}. The simulation scripts are available on Zenodo~\cite{zenodo_singlet_fission}, and
all the figures are available at~\cite{figshare_singlet_fission}.

\begin{acknowledgments}
\noindent
We thank Prof~Jared Cole, Prof~Dane McCamey, and Dr~Murad Tayebjee for insightful discussions. 
We acknowledge funding: by the European Union Horizon Europe research and innovation programme under the Marie Sklodowska-Curie Action for the project SpinSC, the H2020 project EuRyQa, the QuantERA2017 project QuantHEP, the QuantERA2021 project T-NiSQ, and via the Quantum Technology Flagship project PASQuanS2; by the Italian Ministry of University and Research (MUR) via PRIN2022 project TANQU, via the Departments of Excellence grant 2023-2027 Quantum Frontiers, and via the Italian National Centre on HPC, Big Data and Quantum Computing; by the German Federal Ministry of Education and Research (BMBF) via the project QRydDemo; by the World Class Research Infrastructure - Quantum Computing and Simulation Center (QCSC) of Padova University. We acknowledge computational resources by Cineca on the Leonardo machine.
\end{acknowledgments}

\bibliography{main.bbl}

\clearpage
\onecolumngrid

\appendix

\section{Spinful triplet model and magnetic field effects}
\label{a:spinful_model}

The model discussed in the main text can be generalised to account for magnetic field effects by considering all the three local triplet states (\textit{spinful} triplets) $\ket{T_{1,m}}_i$ with $m=-1,0,1$. To do so we introduce a new set of triplet creation operators $\mathcal{T}^\dagger_{i,m}$, that create a triplet with spin quantum number $m$ from the singlet ground state $S_0$ at site $i$.
Together with the singlet operators $\mathcal{S}_i^\dagger$, the triplet operators respect the following relations,
\begin{align}
    & \mathcal{S}_i^\dagger\mathcal{S}_i^\dagger = 0, \\
    & \mathcal{T}_{i,m}^\dagger \mathcal{T}_{i,m'}^\dagger  = 0, \\
    & \mathcal{S}_i^\dagger \mathcal{T}_{i,m}^\dagger = 0.
\end{align}
Furthermore, $[A_i,B_j] = 0$ for every pair of operators acting on different sites. The triplet Hamiltonian $H_\mathrm{T}$ then becomes,
\begin{equation}
    \label{eq:triplet_hamiltonian}
   \begin{split}
     H_T = &\sum_{i=1}^N  \bigg( \varepsilon_T \sum_{m} \Tdagger{i,}{m}\T{i,}{m} + \bm{B}_i\cdot \bm{S}_i + \bm{S}_i^T D \bm{S}_i \bigg)+ \\ & \sum_{(i,j)}\sum_{m,m'} \bigg( J_{T; m,m'} \Tdagger{i,}{m}\T{j,}{m'} + h.c. \bigg) + \\
    &\sum_{(i,j)} \chi_{i,j}^{\mathrm{iso}} \bm{S}_i \cdot\bm{S}_j,
    \end{split} 
\end{equation}
where $\varepsilon_T$ is the triplet energy, $J_{T;m,m'}$ is the $m$-dependent  triplet hopping strength, $\bm{B}_i$ is the magnetic field vector at site $i$, $D$ is the zero-field splitting (ZFS) tensor~\cite{Collins2019}, and $\bm{S}_i = (S_x^{(i)},S_y^{(i)},S_z^{(i)})$ is the vector of spin-1 operators, which can be written in terms of triplet operators $\T{i}{m}$ as follows
\begin{align}
    \label{eq:Sx}
    &S_x^{(i)} = \sum_{m=\pm1}\frac{1}{\sqrt{2}} 
    \bigg(\Tdagger{i,}{0} \T{i,}{m}+h.c.\bigg), \\
    \label{eq:Sy}
    &S_y^{(i)} = \sum_{m=\pm1}\frac{i}{\sqrt{2}} \bigg(m\Tdagger{i,}{0} \T{i,}{m}- m \times h.c.\bigg), \\
    \label{eq:Sz}
    &S_z^{(i)} = \sum_{m=-1}^{1} m\Tdagger{i,}{m} \T{i,}{m}.
\end{align}

The singlet-triplet interaction occurs locally, via coupling between a singlet state $\ket{S_1}_i$ on site $i$, and the triplet-pair state $\ket{{}^1 TT_0}_{ij}$ on sites $i,j$ that has singlet character, i.e., with total spin number $S = 0$~\cite{Miyata2019,Collins2019}. Assuming that the ZFS interaction is a small perturbation of the Zeeman term and the exchange interaction, the triplet-pair state with singlet character is
\begin{equation}
    \label{eq:triplet_pair_singlet}
    \ket{{}^1 TT_0}_{ij} = \frac{1}{\sqrt{3}} \Big( \ket{1,0}_i\ket{1,0}_j - \ket{1,-1}_i \ket{1,1}_j - \\
    \ket{1,1}_i\ket{1,-1}_j\Big),
\end{equation}
which is created by the following combination of triplet creation operators,
\begin{equation}
    \label{eq:TT_singlet_operator}
    \mathcal{T}\hspace{-4.5pt}\mathcal{T}^\dagger_{ij} = \frac{1}{\sqrt{3}}\Big(\Tdagger{0}{i}\Tdagger{0}{j} - \Tdagger{-1}{i}\Tdagger{1}{j} -\Tdagger{1}{i}\Tdagger{-1}{j} \Big)
\end{equation}
This allows us to write the singlet-triplet interaction as
\begin{equation}
    \label{eq:interaction}
        H_\mathrm{int} = \sum_{i=1}^N \gamma \bigg( \mathcal{T}\hspace{-4.5pt}\mathcal{T}^\dagger_{ii+1} \mathcal{S}_i^\pdagger +  \mathcal{T}\hspace{-4.5pt}\mathcal{T}^\dagger_{ii+1}\mathcal{S}_{i+1}^\pdagger + h.c.  \bigg).
\end{equation}

\section{Transition rates from perturbation theory}
\label{a:perturbative_solution}

Singlet fission rates can be calculated perturbatively in the singlet-triplet interaction strength $\gamma$ if the latter is sufficiently small. Here we calculate SF rates $\Gamma$ by treating $H_0 = H_S + H_T$ as the unperturbed Hamiltonian and $H_1 = H_\mathrm{int}$ as the perturbation, for different choices of initial and final states. First, we examine the conserved quantities of the singlet and triplet Hamiltonians in Sec.~\ref{ss:conserved_quantities}, and then we calculate the SF transition elements which are used to calculate SF rates with Fermi's golden rule.

\subsubsection{Conserved quantities and symmetry sectors}
\label{ss:conserved_quantities}

Let us begin by looking at the structure of the unperturbed Hamiltonian $H_0$ and its perturbation $H_1$. The unperturbed Hamiltonian $H_0$ is composed of two commuting terms $[H_\mathrm{S},H_\mathrm{T}] = 0$. Singlet and triplet Hamiltonians also conserve their total number of particles. Therefore, the unperturbed Hamiltonian conserves the total number of both singlet and triplets. We decompose each term in $H_0$ into symmetry sectors associated with a given number $0\leq n\leq N$ of particles in the system
\begin{equation}
    H_0 = \sum_{n=0}^{N} \Big( H_\mathrm{S}^{(n)} + H_\mathrm{T}^{(n)} \Big),
\end{equation}
where $H_\mathrm{S}^{(n)}$ and $H_\mathrm{T}^{(n)}$ are the symmetry sector with $n$ particles of the singlet and triplet Hamiltonians, respectively. Therefore, to calculate SF rates we just need to compute the eigenstates of particular symmetry sectors. Although we mainly focus on the $H_\mathrm{S}^{(1)}$ and $H_\mathrm{T}^{(2)}$ sectors, other combinations can be considered.

\subsubsection{Rates without temperature dependence}
\label{ss:no-temperature}

Let us consider an initial state $\rho = \sum_{\alpha} p_\alpha \ketbra{\varphi_\alpha}{\varphi_\alpha}$ given by a probability distribution over the eigenstates of $H_0$. Then, the Fermi's golden rule transition rates read
\begin{equation}
    \label{eq:FGR}
    \Gamma = \sum_{\alpha\alpha'} p_\alpha \frac{2\pi}{\hbar}|\braket{\varphi_\alpha|H_1|\varphi_{\alpha'}}|^2 \delta(E_\alpha - E_{\alpha'}),
\end{equation}
where $E_\alpha$ is the energy associated with eigenstate $\varphi_\alpha$.
 
This expression is evaluated by calculating the transition amplitudes $|\braket{\varphi_\alpha|H_1|\varphi_{\alpha'}}|^2$ for any combination of initial and final states. Here, we present the form of the transition amplitudes for a few cases of interest.

\paragraph*{1-singlet to 2-triplet:}
Let $\ket{\varphi_\alpha}$ be a 1-singlet state and $\ket{\varphi_{\alpha'}}$ be a 2-triplet state,
\begin{align}
    \label{eq:one_singlet}
    &\ket{\varphi_\alpha} = \sum_{i=1}^N c_i \Sdagger{i}\ket{S_0^{(N)}}, \\
    \label{eq:two_triplet}
    &\ket{\varphi_{\alpha'}} = \sum_{j,k=1}^N \tilde{c}_{jk} \Tdagger{j}{}\Tdagger{k}{}\ket{S_0^{(N)}},
\end{align}
where we have omitted the quantum number $m$ in the triplet operators assuming the there is no external magnetic field and that the ZFS interaction is negligible. Then, the transition elements read
\begin{equation}
    \label{eq:no_field_12_transition_element}    \braket{\varphi_\alpha|H_1|\varphi_{\alpha'}} = g \sum_{i} c_i \tilde{c}_{i,i+1}^*,
\end{equation}
with periodic boundary conditions in $i,i+1$.

In the presence of magnetic fields, we can consider any 2-triplet state given by
\begin{equation}
    \ket{\varphi_{\alpha'}} = \sum_{i=1}^{N} \tilde{c}_{\substack{mm'\\ij}}\Tdagger{i,}{m}\Tdagger{j,}{m'}\ket{S_0^{(N)}},
\end{equation}
to obtain the transition elements
\begin{equation}
    \label{eq:field_12_transition_element}    \braket{\varphi_\alpha|H_1|\varphi_{\alpha'}} = g \sum_{i} c_i \bigg( \tilde{c}_{\substack{00 \\ i,i+1}}^*   - \tilde{c}_{\substack{-+ \\ i,i+1}}^*  - \tilde{c}_{\substack{+- \\ i,i+1}}^*  
    \bigg).
\end{equation}

\paragraph*{Arbitrary transitions at zero magnetic field:}
The transition elements for the case of arbitrary initial and final states can be expressed using the following \textit{Fock-like} notation,
\begin{equation}
    \begin{split}
    \ket{\varphi_\alpha} &=  \sum_{\bm{\alpha}} c_{\bm{\alpha}}\ket{\bm{\alpha}}, \\
    &=\sum_{\alpha_i \cdots\alpha_N} c_{\alpha_1 \cdots \alpha_N} \ket{\alpha_1\cdots \alpha_N},
    \end{split}
\end{equation}
where
\begin{equation}
    \alpha_i = \begin{cases}
        0, \text{if site $i$ is in state $S_0$}, \\
         1, \text{if site $i$ is in state $S_1$}, \\
          2, \text{if site $i$ is in state $T_1$}.
    \end{cases}
\end{equation}
We obtain the following general zero-field transition elements, calculated assuming that the singlet-triplet interaction coefficient is constant for any pair of neighbouring sites,
\begin{equation}
    \label{eq:zero-field_general_transiton_elements}
        \braket{\varphi_\alpha|H_1|\varphi_{\alpha'}} = g \sum_{\bm{\alpha}\bm{\alpha'}} \tilde{c}^*_{\bm{\alpha'}} c_{\bm{\alpha}}^{\phantom{*}} \sum_{\substack{i \\ j\in{\rm NN}(i)}} 
        \bigg( \delta_{\alpha_i,2}\delta_{\alpha_j,2}\delta_{\alpha'_i,1}\delta_{\alpha'_j,0} + \delta_{\alpha_i,1}\delta_{\alpha_j,0}\delta_{\alpha'_i,2}\delta_{\alpha'_j,2} \bigg) 
        \prod_{k\neq i,j}\delta_{\alpha^{\phantom{'}}_k \alpha'_k}.
\end{equation}

\section{Triplet-pair separation}
\label{a:triplet-separation}

The spatial separation of triplets is a key observable that can bring insight into the nature of singlet fission efficiency. Here we quantify triplet-pair separation for the case of $\braket{\mathcal{N}_S}_0 = 1$ (one initial singlet) directly by measuring the expectation value of a ``separation'' operator $Y$
\begin{equation}
    \label{eq:triplet-pair_separation}
    Y = \sum_{i=1}^N\sum_{j=i+1}^N y_{ij} \mathcal{T}^\dagger_i\mathcal{T}^\dagger_j\ket{S_0^{(N)}}\bra{S_0^{(N)}}\mathcal{T}^\pdagger_j\mathcal{T}^\pdagger_i,
\end{equation}
where $y_{ij}:=\mathrm{min}\{|i-j|,N-|i-j|\}$ is the separation between two sites $i$ and $j$ on a 1D ring. For reference $\braket{\psi|Y|\psi} = 0$ for all states $\ket{\psi}$ that do not contain two triplets, and $\braket{\psi|Y|\psi} = 1$ for neighbouring triplet-pair states like $\ket{\psi} = \mathcal{T}^\dagger_i\mathcal{T}^\dagger_{i+1}\ket{S_0^{(N)}}$.

We also evaluate the triplet separation indirectly by monitoring the triplet population on each site, which is useful to visually reveal the presence of spatially separated triplet pairs in the medium.

\begin{figure}[h]
    \centering
    \includegraphics[width=\textwidth]{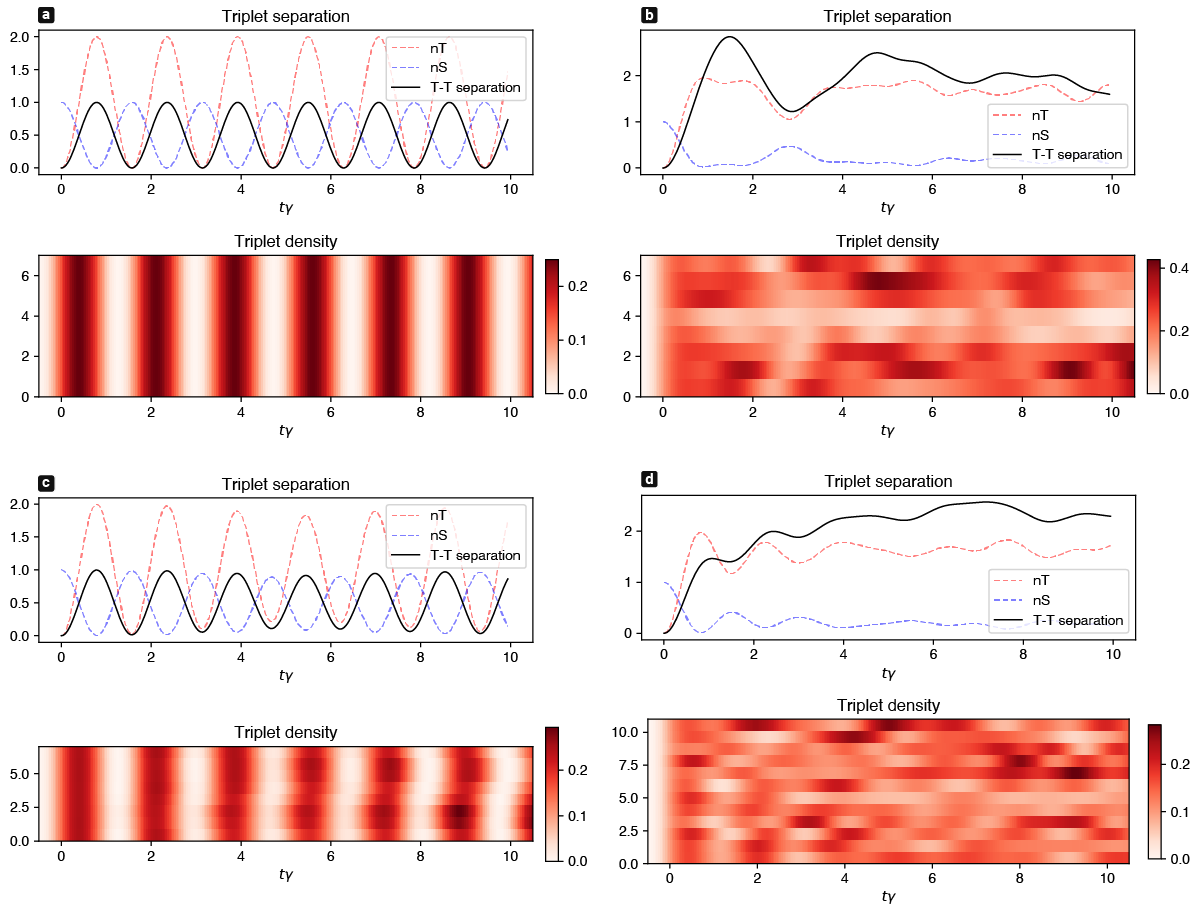}
    \caption{\textbf{Triplet-pair separation.}---Triplet-pair (T-T) separation as a function of time evaluated using Eq.~\eqref{eq:triplet-pair_separation} and the local triplet density (given by local triplet state populations). Solid black lines represent the T-T separation, dashed red and blue lines represent the number of triplets and singlets, respectively. The density plot shows the triplet population at each site (y-axis) over the evolution time. (a) In the resonant triplet-pair solution the T-T separation is limited to 1 (neighbouring sites), here shown for $N=8$. (b) The optimised solution is characterised by large transient T-T separation, which can be larger than 2, here shown for $N=8$. This translates to higher SF efficiency. (c) If the triplet hopping coupling is turned off ($J_T = 0$ from the optimised solution) the triplet separation is again limited to 1, here shown for $N=8$. For larger rings (here $N=12$) the triplet separation can be even larger for longer times. This translates to higher SF efficiency.}
    \label{fig:triplet separation}
\end{figure}

The results, shown in Fig.~\ref{fig:triplet separation} indicate that:
\begin{itemize}
    \item The \textit{resonant triplet-pair solution} is characterised by superpositions of triplet-pair states that are located at neighbouring sites, i.e., $\braket{Y} \leq 1$. See Fig.~\ref{fig:triplet separation} (a).
    \item The optimised solution is characterised by triplet pairs that separate over distant sites, i.e., such that $\braket{Y}$ can be larger than one, as shown in Fig.~\ref{fig:triplet separation} (b). This effect is increasingly evident as the size of the ring increases, as shown in Fig.~\ref{fig:triplet separation} (d).
    \item A non-vanishing triplet hopping strength $J_T > 0$ is necessary to promote triplet separation. By replacing $J_T \to 0$ in the optimised solution the triplet separation is again limited to neighbouring chromophores and the singlet fission efficiency drops to values comparable to those of the resonant triplet-pair solution. See Fig.~\ref{fig:triplet separation} (c).
\end{itemize}

\section{TTN Convergence to analytic solution}
In Sec.~\ref{ss:non-dissipative} and~\ref{ss:dissipative} we presented the results obtained by simulating SF using quantum number conserving TTNs with $U(1)$ symmetry. 
Before simulating the dynamics for arbitrary points in the parameter space we test the TTN model and dynamics against the analytic resonant triplet pair solution of Eq.~\eqref{eq:resonant_solution}. The parameter that defines the accuracy of the method is the bond dimension $\nu_\mathrm{bond}$. The larger the bond dimension, the more accurate the TTN state representation, which is effectively exact when $\nu_\mathrm{bond} = d^{N/2}$, where $d$ is the local dimension of the considered system. 

\begin{figure}[hb]
    \centering
    \includegraphics[width = 0.58\textwidth]{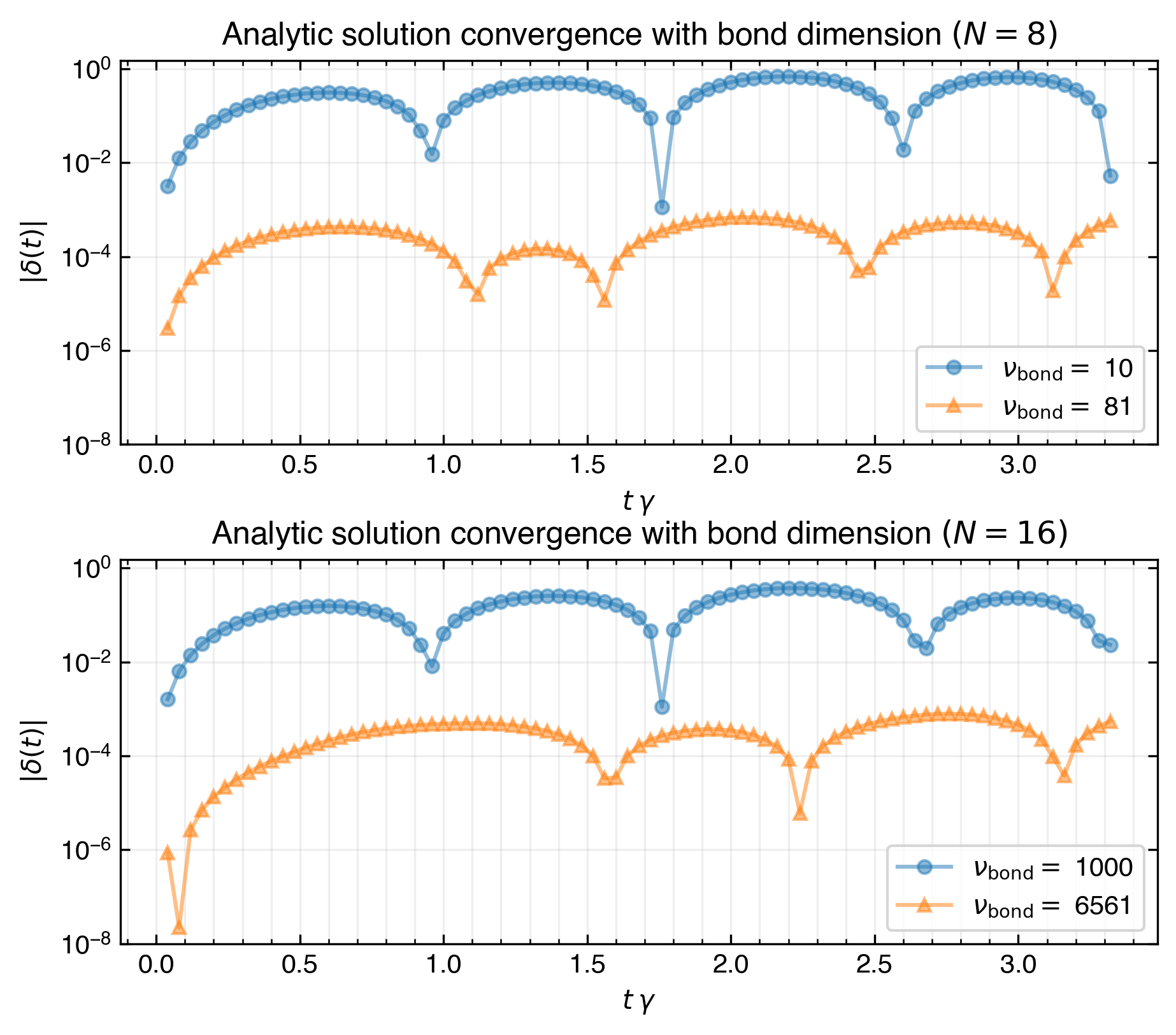}
    \caption{Absolute errors $|\delta(t)|$ between the analytic solution of Eq.~\eqref{eq:resonant_solution} and the TTN results for $N=8$ (top) and $N=16$ (bottom) as a function of the bond dimension $\nu_\mathrm{bond}$, as indicated in the legend. Since convergence can be slow in $\nu_\mathrm{bond}$, we test convergence again for specific points in the parameter space.}
    \label{fig:analytic_convergence}
\end{figure}

In Fig.~\ref{fig:analytic_convergence} we show the absolute error $|\delta(t)| = |\braket{\mathcal{N}_T}_t - \braket{\mathcal{N}_T}_t^{(\nu_\mathrm{bond)}}|$ between the analytic solution of Eq.~\eqref{eq:resonant_solution} and the TTN solution $\braket{\mathcal{N}_T}_t^{(\nu_\mathrm{bond)}}$ with bond dimension $\nu_\mathrm{bond}$ for $N=8,16$. When the maximal bond dimension is reached, the TTN errors are, on average, well below 0.1\%. While errors can be kept within the qualitative range (1-10\%) for lower bond dimensions ($30<\nu_\mathrm{bond}<100$), convergence is slow as indicated for $N=16$. For this reason, we test the convergence of TTN solutions for specific points in the parameter space of the SF Hamiltonian to evaluate the accuracy of the results.

\section{SF efficiency convergence}
\label{a:propagation}

\begin{figure}[hb]
    \centering
    \includegraphics[width = 0.75\textwidth]{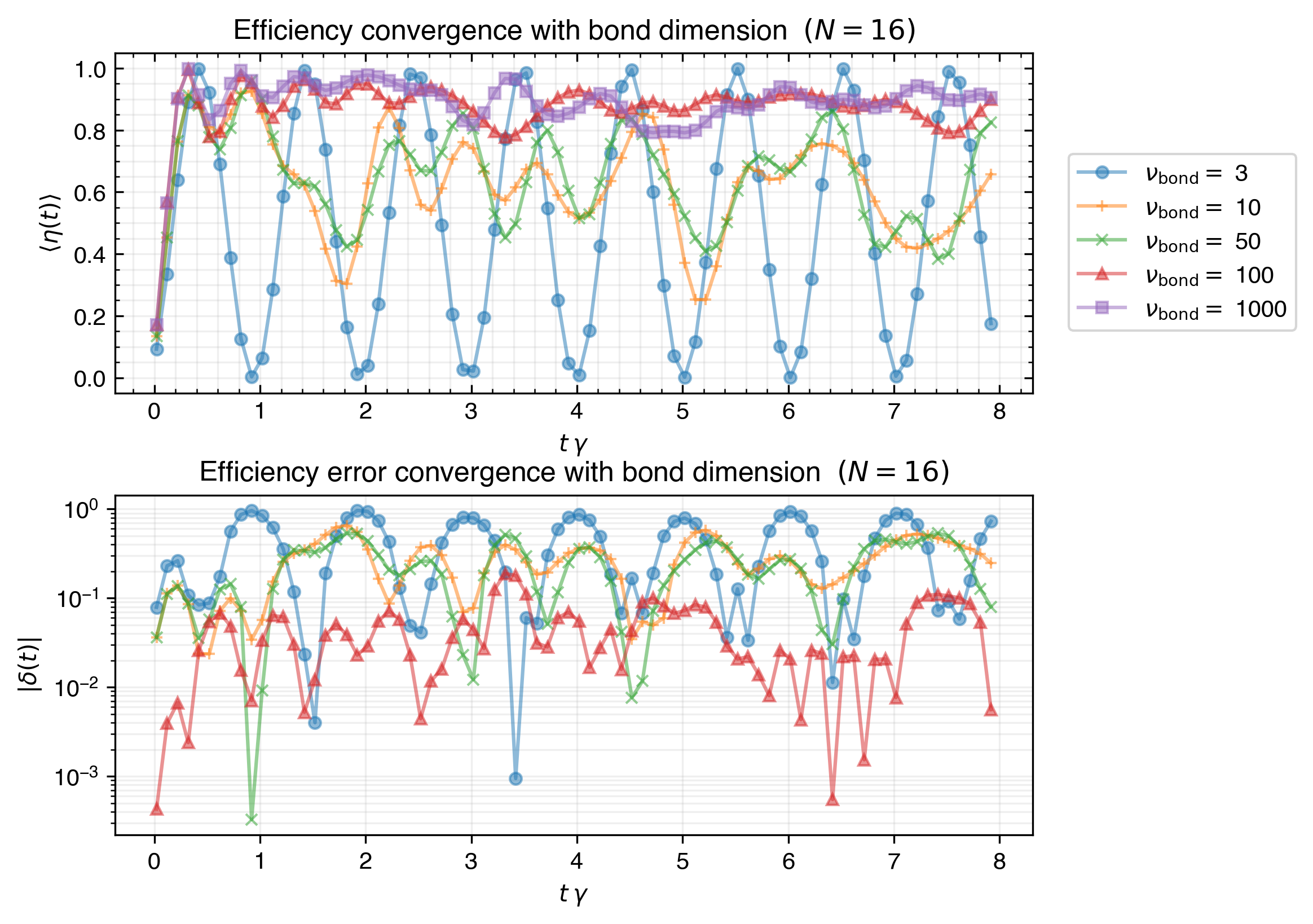}
    \caption{Ensemble-averaged efficiency $\langle\eta(t)^{\nu_\mathrm{bond}}\rangle$ (top) and its error from the solution obtained for $\nu_\mathrm{bond} = 1000$ (bottom) for $N=16$. The accuracy suddenly increases for $\nu_\mathrm{bond}=100$, where it is within 1\% errors from $\nu_\mathrm{bond}=1000$. The results are obtained averaging over 100 disordered realisations. }
    \label{fig:exact_convergence}
\end{figure}

\begin{figure}[hb]
    \centering
    \includegraphics[width = 0.56\textwidth]{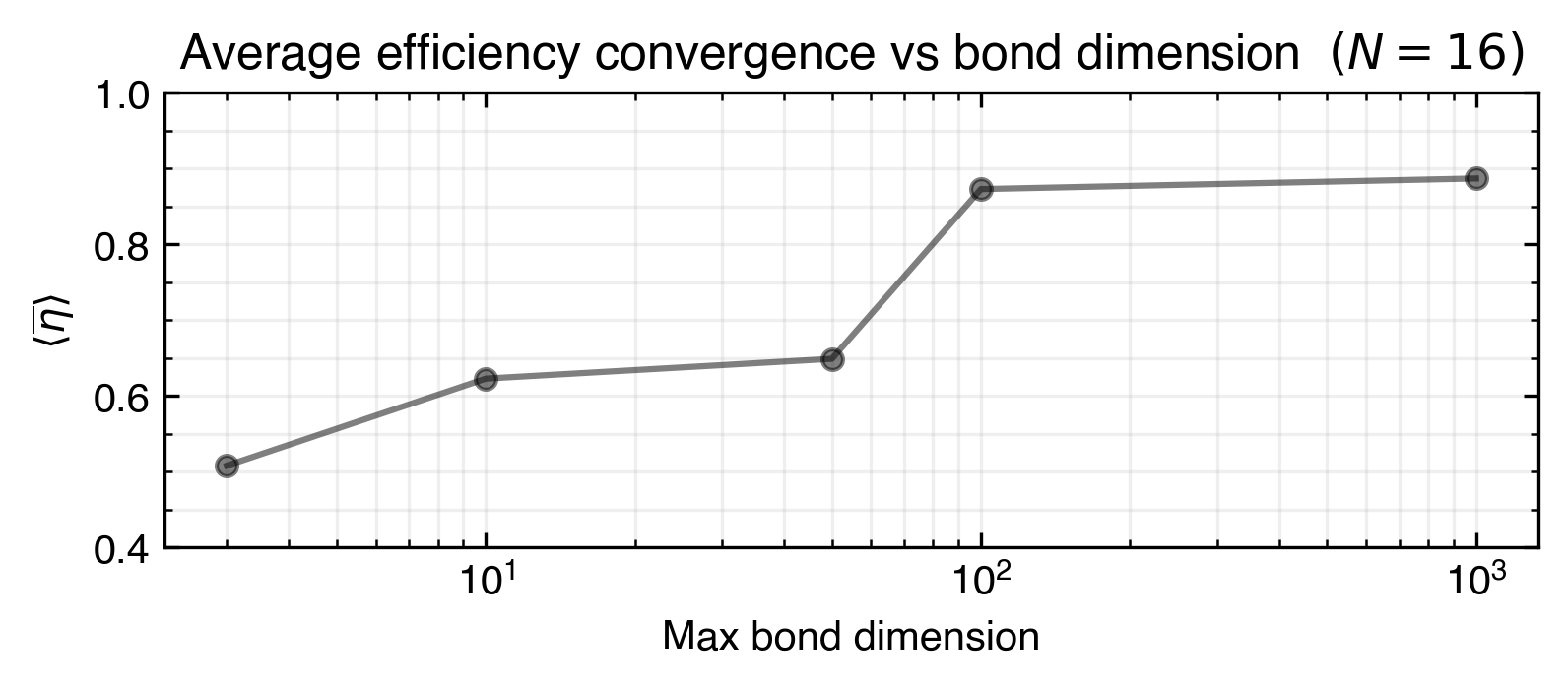}
    \caption{Time-average and ensemble-averaged efficiency $\braket{\overline{\eta}}$ for $N=16$ and $\nu_\mathrm{bond} = 3,10,50,100,1000$. As discussed in Fig.~\ref{fig:exact_convergence}, the efficiency is always underestimated as the bond dimension decreases, and its estimation for $\nu_\mathrm{100}$ is within 1\% from that for $\nu_\mathrm{100}$. The results are obtained averaging over 100 disordered realisations.}
    \label{fig:underestimation}
\end{figure}

To evaluate the efficiency of the solution reported in Tab.~\ref{tab:solutions} for $N\geq 16$ and $n_0\geq 0$ using TTNs, we first test the accuracy of the simulation by calculating the ensemble-averaged efficiency $\braket{\eta(t)}$ for increasing bond dimension $\nu_\mathrm{bond} = 3,10,50,100,1000$. 
Here, $\braket{\eta(t)}$ is the average of $\eta(t)$ over the trajectories obtained for different realisations of disorder.
We also calculate the absolute error $|\delta(t)| = |\braket{\eta(t)^{(\nu_\mathrm{bond})}} - \braket{\eta(t)^{(1000)}}|$ from the solution obtained with $\nu_\mathrm{bond}=1000$. The results, shown in Fig.~\ref{fig:exact_convergence} for $N=16$, indicate that SF efficiency is always underestimated as the bond dimension decreases, meaning that in this region of parameter space TTN simulations offer a lower bound on SF efficiency. For $\nu_\mathrm{bond} = 100$, the efficiency is, on average, within 1\% from the results obtained with $\nu_\mathrm{bond} = 1000$. The effect of underestimation of the efficiency is also shown in Fig.~\ref{fig:underestimation} for $N=16$.

\end{document}

%% file: commands.tex
\definecolor{purple}{HTML}{9326ff}

\def\bra#1{\langle{#1}|}
\def\ket#1{|{#1}\rangle}
\def\braket#1{\langle{#1}\rangle}
\newcommand{\ketbra}[2]{\ket{#1}\!\bra{#2}}

\def\BraVert{\egroup\,\mid\,\bgroup}

\newcommand{\pdagger}{{\phantom{\dagger}}}

\newcommand{\T}[2]{\mathcal{T}_{#1;#2}}